\documentclass[prd,
nofootinbib,
preprintnumbers,
twocolumn,
showpacs
]{revtex4}
\usepackage{amsmath}
\usepackage{amssymb}
\usepackage{epsfig}
\usepackage{graphicx}
\usepackage{longtable}
\newcommand{\capdef}{}
\newcommand{\mycaption}[2][\capdef]{\renewcommand{\capdef}{#2}%
\caption[#1]{{\footnotesize #2}}}
\newcommand{\beq}{\begin{equation}}
\newcommand{\eeq}{\end{equation}}
\newcommand{\bd}{\begin{displaymath}}
\newcommand{\ed}{\end{displaymath}}
\newcommand{\bea}{\begin{eqnarray}}
\newcommand{\eea}{\end{eqnarray}}
\newcommand{\nn}{\nonumber}

\def\eq#1{{Eq.~(\ref{#1})}}

\def\vev#1{\left\langle #1\right\rangle}

\def\ie{\hbox{\it i.e.}{}}
\def\eg{\hbox{\it e.g.}{}}

\newcommand{\mc}[1]{\multicolumn{2}{|c}{#1}}

\begin{document}
%%%%%%%%%%%%%%%%%%%%%%%%%%%%%%%%%%%%%%%%%%%%%%%%%%%%%%%%%%%%%%%%%%%%%
%%%%                     Title-page                              %%%%
%%%%%%%%%%%%%%%%%%%%%%%%%%%%%%%%%%%%%%%%%%%%%%%%%%%%%%%%%%%%%%%%%%%%%
%----------------------------------------------------------------------------------
\title{Fermion masses and mixings in SO(10) models and the neutrino challenge to supersymmetric grand unified theories}
%----------------------------------------------------------------------------------
%\date{\today}
\date{April 28, 2006}
\author{Stefano Bertolini }\email{bertolin@sissa.it}
\affiliation{Scuola Internazionale Superiore di Studi Avanzati,
via Beirut 4, I-34014 Trieste and INFN, Sezione di Trieste, Italy}
\author{Michal Malinsk\'{y}}\email{malinsky@phys.soton.ac.uk}
\affiliation{School of Physics and Astronomy, University of Southampton,
SO16 1BJ Southampton, United Kingdom \\
and CERN Theory Division, CH-1211 Geneva 23, Switzerland.}
\author{Thomas Schwetz}\email{schwetz@sissa.it}
\affiliation{Scuola Internazionale Superiore di Studi Avanzati,
via Beirut 4, I-34014 Trieste and INFN, Sezione di Trieste, Italy}
%----------------------------------------------------------------------------------
\begin{abstract}
We present a detailed study of the quark and lepton mass spectra in a
$SO(10)$ framework with one $10_{H}$ and one $\overline{126}_{H}$ Higgs
representations in the Yukawa sector. We consider in full generality
the interplay between type-I and type-II seesaw for neutrino masses.
We first perform a $\chi^2$ fit of fermion masses independent on the
detailed structure of the GUT Higgs potential and determine the regions of
the parameter space that are preferred by the fermion mass sum rules.
We then apply our study to the case of the minimal renormalizable SUSY
$SO(10)$ GUT with one $10_{H}$, one $\overline{126}_{H}$, one $126_{H}$,
and one ${210}_{H}$ Higgs representations. By requiring that proton
decay bounds are fulfilled we identify a very limited area in the
parameter space where all fermion data are consistently reproduced.
We find that in all cases gauge coupling unification
in the supersymmetric scenario turns out to be severely affected by the presence of
lighter than GUT (albeit $B-L$ conserving) states. We then conclusively
show that the minimal supersymmetric $SO(10)$ scenario here considered is
not consistent with data. The fit of neutrino masses with type-I and
type-II seesaws within a renormalizable $SO(10)$ framework strongly
suggests a non-SUSY scenario for gauge unification.
\vspace*{0ex}
\end{abstract}
%----------------------------------------------------------------------------------
%----------------------------------------------------------------------------------
\maketitle
% the footnote symbols are only redefined for the title page !
\renewcommand{\thefootnote}{\alph{footnote}}
\renewcommand{\thefootnote}{\fnsymbol{footnote}}
\renewcommand{\thefootnote}{\it\alph{footnote}}
\renewcommand{\thefootnote}{\arabic{footnote}}
\setcounter{footnote}{0}

%%%%%%%%%%%%%%%%%%%%%%%%%%%%%%%%%%%%%%%%%%%%%%%%%%%%%%%%%%%%%%%%%
\section{Introduction}
%%%%%%%%%%%%%%%%%%%%%%%%%%%%%%%%%%%%%%%%%%%%%%%%%%%%%%%%%%%%%%%%%

Understanding the pattern of fermion masses and mixings is one of the
longstanding challenges in particle physics. In this respect Grand
Unified Theories (GUTs) do provide an appealing and powerful tool to
address the multiplicity of matter states, by involving stringent
relations among the known fermions (and possibly implying a natural
enlargement of the minimal sector).  Very appealing candidates for a
GUT are models based on the $SO(10)$ gauge group~\cite{GUT}. All the
known fermions plus right-handed neutrinos fit into three
16-dimensional spinorial representations of $SO(10)$, and hence the
model naturally leads to neutrino masses based on the seesaw
mechanism~\cite{seesawI,seesawII}.

In this work we concentrate on a minimal renormalizable version of the
supersymmetric $SO(10)$ GUT, where fermion mass matrices are obtained
from the Yukawa couplings of the matter fields to 10 and 126
dimensional Higgs representations~\cite{Aulakh:1982sw}. The
$\overline{126}_{H}$ representation contains the scalar multiplets $(10,3,1) \oplus
(\overline{10},1,3)$ under the Pati-Salam subgroup~\cite{Pati:1974yy}
${SU}(4)_C\otimes{SU}(2)_L\otimes{SU}(2)_R$, whose vacuum expectation values (VEVs)
generate mass matrices for the left- and right-handed neutrino fields respectively.
Hence, in general there are contributions to the effective mass matrix
of the light neutrinos from the type-I~\cite{seesawI} as well as
type-II~\cite{seesawII} seesaw mechanisms. Thanks to its minimality
the model leads to constrained relations among quark and lepton mass
matrices~\cite{Babu:1992ia,Lavoura:1993vz}:
\begin{eqnarray}
M_d    &=& v_d^{10} Y_{10} +  v_d^{126} Y_{126}\,,\nonumber\\
M_u    &=& v_u^{10} Y_{10} +  v_u^{126} Y_{126}\,,\nonumber\\
M_\ell &=& v_d^{10} Y_{10} -3 v_d^{126} Y_{126}\,, \label{eq:mass-relations}\\
M_D    &=& v_u^{10} Y_{10} -3 v_u^{126} Y_{126}\,,\nonumber\\
M_L    &=& v_L Y_{126} \,, \nonumber \\
M_R    &=& v_R Y_{126} \,. \nonumber
\end{eqnarray}
Here $M_D, M_L, M_R$ denote the Dirac neutrino mass matrix, the mass
matrix of the light left-handed neutrinos, and the mass matrix of the
heavy right-handed neutrinos respectively.  The indices $d,u$ refer
to down- and up-quarks, $\ell$ denotes the charged leptons, while $Y_{10}$ and
$Y_{126}$ are two complex symmetric matrices related to the ${10}_{H}$
and $\overline{126}_{H}$ Yukawa interactions.
The various $v$'s denote the VEVs of the relevant Higgs
multiplets, with $v_{u,d}^{10,126}$ standing for the $SU(2)_L$ doublet
components giving rise to the light MSSM-like Higgses, while $v_R$
($v_L$) are the $SU(2)_L$ singlet (triplet) VEVs entering the
type-I (type-II) seesaw formulae for the light neutrino masses.  The
effective mass matrix for the light neutrinos is then written as
\begin{equation}\label{eq:Mnu}
M_\nu = M_L - M_D M_R^{-1} M_D \,,
\end{equation}
where the symmetry of $M_D$ is used. In Eq.~(\ref{eq:Mnu})
the first and second terms correspond to type-II and type-I seesaws
respectively. Hence, all fermion mass matrices are predicted in terms of
two complex symmetric Yukawa matrices and six VEVs.

Quite an amount of activity has been devoted recently to the study of
the minimal renormalizable SUSY $SO(10)$ model, see for instance
Refs.~\cite{Aulakh:2000sn,Fukuyama:2002ch,Bajc:2002iw,Goh:2003sy,Fukuyama:2003hn,Aulakh:2003kg,Goh:2003hf,Bajc:2004xe,Bajc:2004fj,Dutta:2004wv,Goh:2004fy,Aulakh:2002zr,Fukuyama:2004ti,Aulakh:2005ic,Fukuyama:2005us,Bertolini:2004eq,Bertolini:2005qb,Babu:2005ia,Dutta:2005ni}.
This interest has been triggered to some extent by the experimental
progress in neutrino physics, and the observation of
Refs.~\cite{Bajc:2002iw,Goh:2003sy}, that within the minimal $SO(10)$
model with type-II seesaw dominance $b-\tau$ unification allows naturally for a
large lepton mixing due to the
destructive interference between the 33 elements of the down-quark and
charged lepton mass matrices. Analyses of proton decay within this
setting are found, \eg, in Refs.~\cite{Goh:2003nv,Fukuyama:2004pb},
while various extensions of the model have been considered in view
of rising some of the tensions with the data, by adding non-renormalizable terms~\cite{Dutta:2004wv},
by minimally extending the Higgs sector~\cite{Goh:2004fy},
or by including Yukawa couplings of the fermion fields with an additional
120-dimensional Higgs representation~\cite{Bertolini:2004eq,Bertolini:2005qb,Dutta:2005ni}.

Most of the previously quoted analyses consider the dominance of one
type of seesaw and assume that the vacuum gives the correct neutrino
mass scale. Very recently two studies appeared showing an intrinsic
antagonism between seesaw neutrino masses and coupling unification in
the minimal SUSY $SO(10)$ scenario~\cite{Bajc:2005qe,Aulakh:2005mw},
pointing out a critical tension between the needed seesaw neutrino
scale and the detailed spectrum arising from the minimal $SO(10)$ breaking GUT
vacuum~\cite{Bajc:2004xe,Aulakh:2002zr,Aulakh:2005bd}.

The aim of the present paper is twofold.  First, in
Secs.~\ref{sec:analysis} and \ref{sec:results} we study in detail the
implications of the fermion mass sum rules that emerge from $10_H$ and
$\overline{126}_H$ Yukawa terms in a $SO(10)$ GUT framework. The
analysis is performed in full generality, allowing for complex Yukawas
and VEVs, and considering the neutrino mass matrix as originating from
an admixture of type-I and type-II seesaw as given by
Eq.~(\ref{eq:Mnu}). In the second part of the paper,
Sec.~\ref{sec:minimalso10}, we specialize our study to the minimal
renormalizable SUSY $SO(10)$ model, by including the
constraints coming from the detailed structure of the GUT symmetry
breaking vacuum and performing an exhaustive study of the parameter
space by optimizing a $\chi^2$ function.
This approach is complementary to the random parameter searches applied in
previous studies, since the algorithm converges to an optimal solution
(if it exists), and isolated solutions (allowed by an acceptable
fine tuning of the parameters) can be found, which are easily missed
in a random parameter scan.

We show conclusively that the minimal renormalizable $SO(10)$
supersymmetric scenario does not allow for a consistent fit of the
fermion mass spectrum, the model being unable to reproduce the correct
neutrino mass scale via type-I and/or type-II seesaw. Our negative
result can be cast in more general (and simple) terms considering
that, while i) the present constraints on proton decay force any GUT
scale to lie above $10^{16}$~GeV and ii) SUSY gauge unification works
very well considering a desert scenario between the weak and the GUT
scales, the neutrino mass coming from type-I and/or type-II seesaws,
being proportional via Yukawa and Higgs potential couplings to
$m_{weak}^2/M$, requires for $m_{weak}\approx 200$~GeV, $M <
10^{15}$~GeV. As a consequence (barring a strong interacting sector
in the theory), intermediate mass scales appear which are bound
to affect SUSY gauge coupling unification at the least.

All this applies to the minimal renormalizable
setup. As it was pointed out very recently~\cite{Aulakh:2006vi},
non-minimal realizations of the renormalizable Yukawa sector such as
those containing an additional 120-dimensional Higgs multiplet may
provide a way out of the issues: the neutrino mass scale can be
enhanced by the type-I seesaw with right-handed neutrino masses below
the GUT scale due to tiny Yukawa couplings associated to the
$\overline{126}_{H}$ multiplet, whose role in the charged matter
sector can be taken over by the new $120_{H}$ multiplet. Although the
simplest realization of this programme (where the charged matter
sector is dominated by $10_{H}$ and $120_{H}$ and the Yukawas of
$\overline{126}_{H}$ are neglected) seems to fail
\cite{Lavoura:2006dv}, this attempt does provide a direction
for further studies. Alternatively, Planck induced non-renormalizable
operators may also provide the scale suppression needed by the
neutrino sector, see for instance Ref.~\cite{Mohapatra:2006gs} and
references therein.

On the other hand, the simplicity and the high level of predictivity
that makes the minimal renormalizable SUSY $SO(10)$ GUT extremely
appealing is lost in both extensions. Before calling the minimal
setup a dead end an extensive analysis of potential loopholes in the
abovementioned arguments is due. This aim we pursue with the present
paper up to, helas, the bitter end.

%%%%%%%%%%%%%%%%%%%%%%%%%%%%%%%%%%%%%%%%%%%%%%%%%%%%%%%%%%%%%%%%
\section{Description of the analysis}
%%%%%%%%%%%%%%%%%%%%%%%%%%%%%%%%%%%%%%%%%%%%%%%%%%%%%%%%%%%%%%%%
\label{sec:analysis}

\subsection{The parameterization}

In order to investigate whether Eqs.~(\ref{eq:mass-relations}) and
(\ref{eq:Mnu}) allow for fermion masses and mixing in agreement with
the data we proceed as follows. It turns out to be convenient to
express the $Y_{10}$ and $Y_{126}$ Yukawa matrices in terms of
$M_\ell$ and $M_d$, and substitute them in the expressions for
$M_u,M_D$ and $M_\nu$:
\begin{eqnarray}
M_u &=& f_u \left[  (3+r)M_d +  (1-r) M_\ell \right] \,,
\label{eq:Mu}\\[1ex]
M_D &=& f_u \left[ 3(1-r)M_d + (1+3r) M_\ell \right] \,,
\label{eq:MD}
\end{eqnarray}
where
\begin{equation}
f_u = \frac{1}{4} \, \frac{v_u^{10}}{v_d^{10}} \,,\qquad
r = \frac{v_d^{10}}{v_u^{10}} \, \frac{v_u^{126}}{v_d^{126}} \,.
\label{eq:fu-r}
\end{equation}
The neutrino mass matrix is obtained as
\begin{equation}
M_\nu = f_\nu \left[ (M_d - M_\ell) +
\xi \, \frac{M_D}{f_u}(M_d-M_\ell)^{-1}\frac{M_D}{f_u} \right] \,,
\label{eq:Mnu2}
\end{equation}
with
\begin{equation}
f_\nu = \frac{1}{4} \, \frac{v_L}{v_d^{126}} \,,\qquad
\xi = - \frac{\left(4 f_u v_d^{126}\right)^2}{v_L v_R} \,.
\label{eq:fnu-xi}
\end{equation}
The parameter $|\xi|$ controls the relative importance of the type-I
and type-II seesaw terms: For $|\xi| \to 0$ one obtains pure type-II
seesaw, whereas $|\xi| \to \infty$ (with $f_\nu |\xi| \simeq$~const.)
corresponds to type-I seesaw. For $|\xi| \sim 1$ both contributions
are comparable.

In what follows we denote diagonal mass matrices by $\hat m_x$,
$x=u,d,\ell,\nu$, with eigenvalues corresponding to the
particle masses, \ie, being real and positive.
We choose a basis where the down-quark matrix is diagonal: $M_d =
\hat m_d$. In this basis $M_\ell$ is a general complex symmetric
matrix, that can be written as $M_\ell = W_\ell^\dagger \hat m_\ell
W_\ell^*$, where $W_\ell$ is a general unitary matrix. Without loss of
generality $f_u$ and $f_\nu$ can be taken to be real and
positive. Hence, the independent parameters are given by 3 down-quark
masses, 3 charged lepton masses, 3 angles and 6 phases in $W_\ell$,
$f_u, f_\nu$, together with two complex parameters $r$ and $\xi$:
21 real parameters in total, among which 8 phases.
Using Eqs.~(\ref{eq:Mu}), (\ref{eq:MD}), and (\ref{eq:Mnu2}) all observables
(6 quark masses, 3 CKM angles, 1 CKM phase, 3 charged
lepton masses, 2 neutrino mass-squared differences, the mass of the
lightest neutrino, and 3 PMNS angles, 19 quantities altogether)
can be calculated in terms of these input parameters.
Although the number of parameters is larger than the number of
observabels the system is sensibly over-constrained due to the non-linear
structure of the problem.

Since we work in a basis where the down-quark mass matrix is diagonal
the CKM matrix is given by the unitary matrix diagonalizing the
up-quark mass matrix up to diagonal phase matrices:
\begin{equation}
\hat m_u = W_u M_u W_u^T
\end{equation}
with
\begin{equation}\label{eq:Vckm}
W_u = \mathrm{diag}(e^{i\beta_1},
e^{i\beta_2}, e^{i\beta_3}) \, V_\mathrm{CKM} \,
\mathrm{diag}(e^{i\alpha_1}, e^{i\alpha_2}, 1) \,,
\end{equation}
where $\alpha_i,\beta_i$ are unobservable phases at low energy. The neutrino mass
matrix given in Eq.~(\ref{eq:Mnu2}) is diagonalized by $\hat
m_\nu = W_\nu M_\nu W_\nu^T$, and the PMNS matrix is determined by
$W_\ell^* W_\nu^T = \hat D_1 V_\mathrm{PMNS} \hat D_2$, where $\hat D_1$
and $\hat D_2$ are diagonal phase matrices similar to those in ~\eq{eq:Vckm}.

\subsection{Input data and $\chi^2$ analysis}

\begin{table}
\centering
  \begin{tabular}{lr}
  \hline\hline
  observable & input data \\
  \hline
  $m_d$ [MeV]   & $1.24 \pm 0.41$\\
  $m_s$ [MeV]   & $21.7 \pm 5.2$\\
  $m_b$ [GeV]   & $1.06^{+0.14}_{-0.09}$\\[1ex]
  $m_u$ [MeV]   & $0.55 \pm 0.25$\\
  $m_c$ [MeV]   & $210 \pm 0.21$\\   % this is the error used in the fit
  $m_t$ [GeV]   & $82.4^{+30.3}_{-14.8}$\\[1ex]
  $\sin\phi^\mathrm{CKM}_{23}$ & $0.0351\pm 0.0013$\\
  $\sin\phi^\mathrm{CKM}_{13}$ & $0.0032\pm 0.0005$\\
  $\sin\phi^\mathrm{CKM}_{12}$ & $0.2243\pm 0.0016$\\
  $\delta_\mathrm{CKM}$        & $60^\circ \pm 14^\circ$\\[1ex]
  $\sin^2\theta^\mathrm{PMNS}_{23}$ & $0.50 \pm 0.065$\\
  $\sin^2\theta^\mathrm{PMNS}_{13}$ & $ < 0.0155$ \\
  $\sin^2\theta^\mathrm{PMNS}_{12}$ & $0.31\pm 0.025$\\[1ex]
  $\Delta m^2_{21}$ [eV$^2$]        & $(7.9 \pm 0.3)\, 10^{-5}$\\
  $|\Delta m^2_{31}|$ [eV$^2$]        & $(2.2^{+0.37}_{-0.27})\, 10^{-3}$\\
  \hline\hline
  \end{tabular}
  \mycaption{Sample of GUT scale input data used in this work (central values
  and 1$\sigma$ errors) for $M_\mathrm{SUSY} = 1$~TeV, $\tan\beta =
  10$, and $M_\mathrm{GUT} = 2\times 10^{16}$~GeV (see text for
  details and references). Charged lepton masses are listed in
  \eq{eq:ch_lept}.
\label{tab:data}}
\end{table}

As input data we use quark and lepton masses and mixing angles
evaluated at the GUT scale, based on the RGE analysis of
Ref.~\cite{Das:2000uk}. As a typical example we consider a SUSY scale
$M_\mathrm{SUSY} = 1$~TeV, $\tan\beta = 10$, and a GUT scale
$M_\mathrm{GUT} = 2\times 10^{16}$~GeV.
Since charged lepton masses are known with an accuracy of better than $10^{-3}$
we do not consider them as free parameters but fix them to
the (GUT scale) central values from Ref.~\cite{Das:2000uk}:
\begin{eqnarray}
\label{eq:ch_lept}
m_e & = & 0.3585\,\mathrm{MeV}\,,\nn \\
m_\mu & = & 75.67\,\mathrm{MeV}\,,\\
m_\tau & =&  1292.2\,\mathrm{MeV}\,.\nn
\end{eqnarray}
The remaining data are listed in Tab.~\ref{tab:data}.
For the heavy quark masses $m_c,m_b,m_t$ we adopt the values
and uncertainties used in Ref.~\cite{Das:2000uk}, whereas for the light quarks
we update the data to the ranges given in Ref.~\cite{PDG}: $m_u
= 1.5 - 4$~MeV, $m_d = 4 - 8$~MeV, $m_s = 80 - 130$~MeV
($\overline{\mathrm{MS}}$ masses at 2~GeV).
The corresponding values at the GUT scale are given in Tab.~\ref{tab:data},
together with the CKM angles and phase~\cite{Bertolini:2005qb}.

Concerning the neutrino parameters, we note that in the setup under
consideration the neutrino mass spectrum is generally normal and
hierarchical with $m_1 < m_2 < m_3$. In this case the RGE running of
the PMNS angles~\cite{Antusch:2003kp,Antusch:2005gp} is of order
$10^{-5}(1+\tan^2\beta) \sim 10^{-3}$, and therefore negligible to a good
approximation. The running of the neutrino masses
is small as well (a running effect common to all three
neutrino masses can be absorbed in the free parameter $f_\nu$, compare
~\eq{eq:Mnu2}). In our analysis we use the
low energy neutrino parameters, as obtained from recent global fits
to neutrino oscillation data~\cite{Maltoni:2004ei,Schwetz:2005jr}.
We do not include any constraint on the lightest neutrino mass $m_1$,
since the values that are obtained are much
below the sensitivity of the present data.

Let us denote the central values and errors of the observables by
$O_i$ and $\sigma_i$, where $i = 1, \ldots, 15$ runs over all the
quantities listed in Tab.~\ref{tab:data}. As described above, the
predictions of these observables, $P_i$, depend on the parameters
$x_\alpha$, where $\alpha = 1, \ldots, 18$ runs over 3 down-quark
masses, 9 real parameters in $W_\ell$, $f_u$, $f_\nu$, $|r|$,
arg($r$), $|\xi|$, arg($\xi$).\footnote{Since we fix the charged
lepton masses to the values given in Eq.~(\ref{eq:ch_lept}) they are
neither included in the set of observables $O_i$, nor among the parameters
$x_\alpha$.} Then a $\chi^2$-function is constructed as
\begin{equation}\label{eq:chisq}
\chi^2(x_\alpha) = \sum_{i=1}^{15}
\left( \frac{P_i(x_\alpha) - O_i}{\sigma_i} \right)^2 \,.
\end{equation}
The data are fitted by minimizing this function with respect to the
parameters $x_\alpha$. The minimization of the scale factors $f_u$ and
$f_\nu$ can be done analytically. The remaining 16 dimensional
minimization is performed with an algorithm based on the so-called
down-hill simplex method~\cite{NR}.

Let us note that the minimization is technically rather challenging.
The problem involves parameters which differ by many orders of
magnitude, and some of the solutions are extremely fine-tuned, which
leads to very steep valleys in the $\chi^2$ landscape. Moreover, due
to the high dimensionality and the non-linearity of the problem there
is a large number of local minima. In the numerical analysis much
effort has been devoted to find the absolute minimum, involving random
methods and dedicated investigations to each particular
problem under consideration.
However, as it is well known, by numerical methods it is difficult to
assess with absolute confidence that an absolute minimum has been found.
Although each minimum has been carefully tested
for possible improvements one should always keep in mind the
possibility that a better solution might exist somewhere in the
parameter space.

%%%%%%%%%%%%%%%%%%%%%%%%%%%%%%%%%%%%%%%%%%%%%%%%%%%%%%%%%%%%%%%%%%%%
\section{General fit of fermion masses and mixings}
%%%%%%%%%%%%%%%%%%%%%%%%%%%%%%%%%%%%%%%%%%%%%%%%%%%%%%%%%%%%%%%%%%%%
\label{sec:results}

\begin{figure}[t]
\centering
\includegraphics[width=0.48\textwidth]{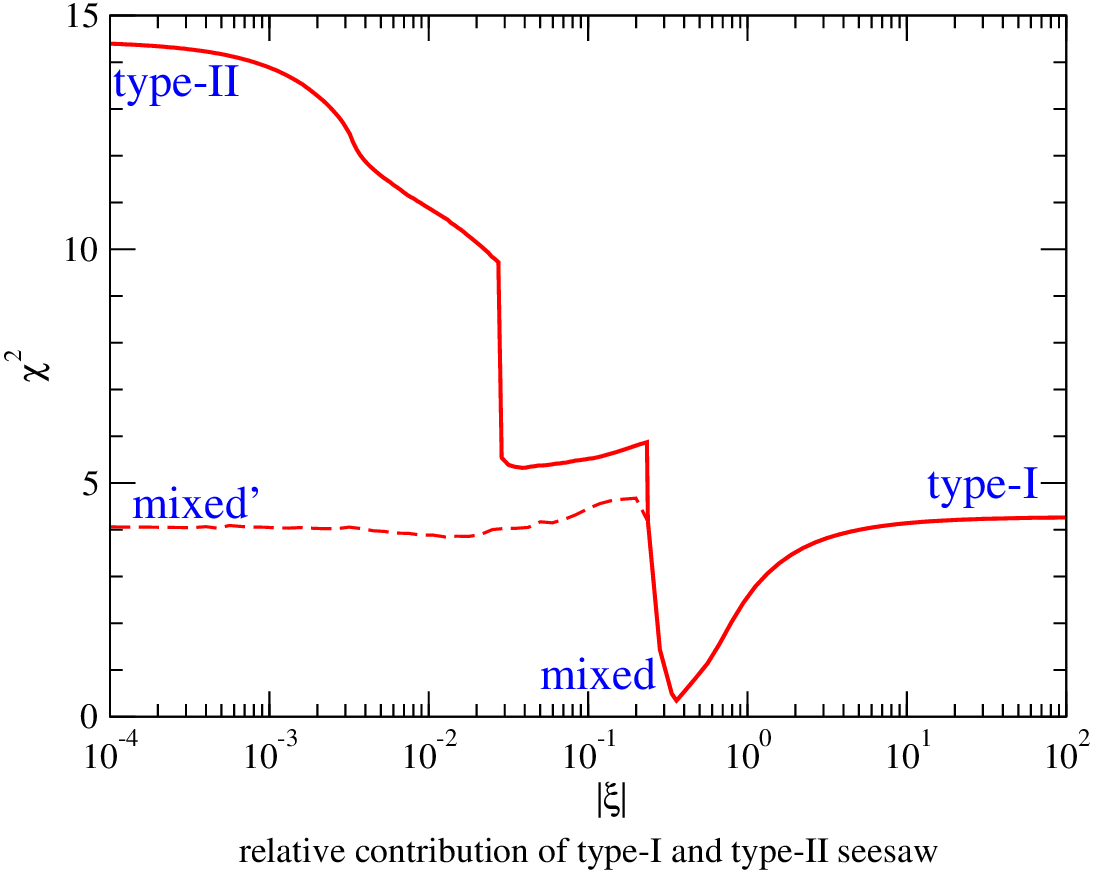}
%-eps-includegraphics[width=0.48\textwidth]{figs/xi-chisq.eps}
%
  \mycaption{$\chi^2$ as a function of the parameter $|\xi|$ which
  controls the relative relevance of type-I and type-II seesaw
  terms. The dashed curve corresponds to the singular (fine tuned) solution
  denoted by {\it mixed'} (see text).}
\label{fig:xi}
\end{figure}

For the discussion of the obtained fits we classify the solutions by
the values of the parameter $|\xi|$ which controls the relative weight
of type-I and type-II seesaw mechanism. In Fig.~\ref{fig:xi} the
$\chi^2$ minimum is shown as a function of this parameter. This means
that for fixed $|\xi|$, $\chi^2(x_\alpha)$ is minimized with respect
to all the other parameters $x_\alpha$ with $\alpha \neq |\xi|$. In
Tab.~\ref{tab:solutions} some parameter values and the predictions for
the data are given for four sample points.

One of the main results of this work is clearly visible in
Fig.~\ref{fig:xi}: We find a pronounced minimum for $|\xi|\simeq 0.36$
which corresponds to a mixture of type-I and type-II seesaw with a
comparable size of both terms. Such a ``mixed'' solution provides an
excellent fit to the data with $\chi^2 \approx 0.35$. From the results
given in Tab.~\ref{tab:solutions} in the column labeled ``mixed'' it
is seen that all observables are fitted within $\lesssim
0.4\sigma$. In particular, all the lepton mixing angles and the
neutrino mass-squared differences are very close to their experimental
values. Also CKM CP-violation is described correctly by the value
$\delta_\mathrm{CKM} = 61^\circ$ obtained in this solution. We
conclude that a scenario with both seesaw terms of comparable size
allows for an excellent description of fermion masses and mixings,
confirming the results of Ref.~\cite{Babu:2005ia}.
Whether this solution is viable still depends on the detailed study
of the global vacuum of the given $SO(10)$ model.

\begin{table*}
%\begin{table}[t]
\renewcommand{\arraystretch}{0.6}
%\begin{longtable}{l|rr|rr|rr|rr}
\begin{tabular}{l|rr|rr|rr|rr}
  \hline\hline
             & \mc{\bf type-II} & \mc{\bf mixed'}& \mc{\bf mixed} & \mc{\bf type-I} \\
  \hline
  $|\xi|$    & \mc{$0$}                   &\mc{$10^{-4}$}             &\mc{$0.3587$}              &\mc{$3.59\times 10^6$}\\
  arg($\xi$) & \mc{$-$}           &\mc{$0.866\pi$}            &\mc{$1.018\pi$}            &\mc{$1.318\pi$}\\
  $|r|$      & \mc{$0.3278$}          &\mc{$1.9977$}              &\mc{$0.47896$}             &\mc{$0.3551$}\\
  arg($r$)   & \mc{$0.408\pi$}        &\mc{$1.849\pi$}            &\mc{$0.0013\pi$}           &\mc{$0.0057\pi$}\\
  $f_u$      & \mc{$16.62$}       &\mc{$11.51$}               &\mc{$18.77$}               &\mc{$19.23$}\\
  $f_\nu$    & \mc{$1.671\times 10^{-10}$}&\mc{$4.519\times 10^{-10}$}&\mc{$8.732\times 10^{-10}$}&\mc{$3.613\times 10^{-17}$}\\
  \hline
  observable & pred. & pull & pred. & pull & pred. & pull & pred. & pull \\
  \hline
  $m_d$ [MeV]                       &$0.7662 $&$\bf-1.16$ & $ 0.4956$ & $\bf-1.82$& $  1.122$ & $-0.29$& $   0.4719$ & $\bf-1.87$\\
  $m_s$ [MeV]                       &$31.33  $&$\bf 1.85$ & $  22.46$ & $    0.15$& $  22.85$ & $ 0.22$& $    19.99$ & $   -0.33$\\
  $m_b$ [MeV]                       &$1147   $&$    0.61$ & $   1096$ & $    0.25$& $   1078$ & $ 0.13$& $     1029$ & $   -0.35$\\[1ex]
  $m_u$ [MeV]                       &$0.5543 $&$    0.02$ & $ 0.5576$ & $    0.03$& $ 0.5512$ & $ 0.00$& $   0.5538$ & $    0.02$\\
  $m_c$ [MeV]                       &$213.1  $&$    0.17$ & $  213.5$ & $    0.18$& $  210.6$ & $ 0.03$& $    213.1$ & $    0.16$\\
  $m_t$ [MeV]                       &$78030  $&$   -0.29$ & $  77411$ & $   -0.34$& $  81659$ & $-0.05$& $    78117$ & $   -0.29$\\[1ex]
  $\sin\phi^\mathrm{CKM}_{23}$      &$0.0345 $&$   -0.43$ & $ 0.0352$ & $    0.08$& $ 0.0351$ & $ 0.03$& $   0.0349$ & $   -0.13$\\
  $\sin\phi^\mathrm{CKM}_{13}$      &$0.00331$&$    0.23$ & $0.00319$ & $   -0.02$& $0.00319$ & $-0.01$& $  0.00323$ & $    0.06$\\
  $\sin\phi^\mathrm{CKM}_{12}$      &$0.2245 $&$    0.11$ & $ 0.2243$ & $    0.02$& $ 0.2243$ & $ 0.01$& $   0.2243$ & $    0.01$\\
  $\delta_\mathrm{CKM}\, [^\circ]$  &$79.35  $&$\bf 1.38$ & $  59.47$ & $   -0.04$& $  61.41$ & $ 0.10$& $    61.11$ & $    0.08$\\[1ex]
  $\sin^2\theta^\mathrm{PMNS}_{23}$ &$0.3586 $&$\bf-2.17$ & $ 0.5126$ & $    0.19$& $ 0.5027$ & $ 0.04$& $   0.4944$ & $   -0.09$\\
  $\sin^2\theta^\mathrm{PMNS}_{13}$ &$0.0145 $&$    0.93$ & $ 0.0106$ & $    0.68$& $ 0.0066$ & $ 0.43$& $   0.0095$ & $    0.61$\\
  $\sin^2\theta^\mathrm{PMNS}_{12}$ &$0.2829 $&$\bf-1.08$ & $ 0.3078$ & $   -0.09$& $ 0.3094$ & $-0.02$& $   0.3078$ & $   -0.09$\\[1ex]
  $\Delta m^2_{21}$[$10^{-5}$eV$^2$]&$7.863  $&$   -0.12$ & $  7.894$ & $   -0.02$& $  7.898$ & $-0.01$& $    7.896$ & $   -0.01$\\
  $\Delta m^2_{31}$[$10^{-3}$eV$^2$]&$2.385  $&$    0.50$ & $  2.232$ & $    0.09$& $  2.210$ & $ 0.03$& $    2.223$ & $    0.06$\\
  $m_1 / \sqrt{\Delta m^2_{21}}$    &$0.279  $&$        $ & $  0.478$ & $        $& $  0.382$ & $     $& $    0.361$ & $        $\\[1ex]
  $\delta_\mathrm{PMNS}\, [^\circ]$ &$-0.70  $&$   	$ & $-59    $ & $   	 $& $-0.70  $ & $     $& $4.9      $ & $    	$\\
  $\alpha_{1}\, [^\circ]$ 	    &$1.1    $&$   	$ & $30     $ & $   	 $& $1.8    $ & $     $& $-2.1     $ & $    	$\\
  $\alpha_{2}\, [^\circ]$ 	    &$91     $&$   	$ & $126    $ & $   	 $& $-84    $ & $     $& $90       $ & $    	$\\
  \hline
  $\chi^2$ & & $14.5$ & & $4.1$ & & $0.35$ & & $4.3$ \\
  \hline\hline
%\end{longtable}
 \end{tabular}
  \mycaption{Parameter values and predictions in four example solutions
  corresponding to different terms dominating the neutrino mass
  matrix: type-I, type-II, or both contributions of comparable size
  (mixed and mixed'). In the column ``pred.'' the predicted
  values $P_i$ for the observables are given, the column ``pull''
  shows the number of standard deviations from the observations, $(P_i
  - O_i)/\sigma_i$, using the data and errors from
  Tab.~\ref{tab:data}. Deviations of more than $1\sigma$ are
  highlighted in boldface. The final $\chi^2$ is the sum of the
  squares of the numbers in the ``pull'' column. See the text for comments on the
values of the leptonic CP phases.}
 \label{tab:solutions}
\end{table*}

The $\chi^2$ increases by increasing $|\xi|$, and it approaches a
value of $\chi^2 \simeq 4.3$ for $|\xi| \gtrsim 10$, when the neutrino
mass matrix becomes dominated by the type-I seesaw term. Also in the
case of complete type-I dominance the fit is very good, with most
observables within $\lesssim 0.3\sigma$, with the sole exception of
the down-quark mass $m_d$ which shows a $-1.87\sigma$ deviation from
its prediction (see Tab.~\ref{tab:solutions}, column ``type-I''). As
we will discuss in more detail later, a low value of $m_d$ is required
for a valuable type-I seesaw fit. Let us note that all neutrino
parameters are in excellent agreement with the observations, and in
particular, the correlation between $\theta_{12}^\mathrm{PMNS}$,
$\theta_{23}^\mathrm{PMNS}$, and the ratio $\Delta m^2_{31}/\Delta
m^2_{21}$ found in Ref.~\cite{Babu:2005ia} seems not to apply here.
This solution is not plagued by the problem of accomodating the correct
$\delta_\mathrm{CKM}$ phase found in Ref.~\cite{Dutta:2004wv},
and discussed on general grounds in Ref.~\cite{Bertolini:2005qb}.

For a pure type-II solution with $|\xi| = 0$ there is more
tension in the fit. Although a $\chi^2 \simeq 14.5$ might be
acceptable for 15 data points from a statistical point of view,
several observables are $1$ to $2 \sigma$ away from the central value.
Among all, one needs a large value of the strange-quark mass,
namely $m_s \simeq 31$~MeV that is $1.85\sigma$ too large, while the PMNS
angle $\theta_{23}$ is too small by $2.2\sigma$. Furthermore, $m_d$,
the CKM phase, and the PMNS angle $\theta_{12}$ show a pull greater than
$1\sigma$. These results agree with previous analyses of pure type-II
solutions~\cite{Bertolini:2005qb,Babu:2005ia,Goh:2003hf}.

The dashed line in Fig.~\ref{fig:xi} corresponds to an interesting
variant of a solution with comparable type-I and type-II
contributions. Formally this solution has a rather small value of
$|\xi|$, which would signal type-II dominance. However, in this case
one eigenvalue of  $(M_d - M_\ell)$ is very small, \ie, this matrix
is close to  singular, which implies that its inverse
has large entries. As a consequence the second term in
Eq.~(\ref{eq:Mnu2}) turns out to be comparable to the first term,
in spite of the small value of $|\xi|$.  Let us denote the type-II term in
Eq.~(\ref{eq:Mnu2}) by $M^\mathrm{II}$ and the type-I term by
$M^\mathrm{I}$.
Then for the solution denoted {\it mixed'} in Fig.~\ref{fig:xi}
and Tab.~\ref{tab:solutions} we find that the matrix entries are of a
similar size: $ 0.2 \lesssim |M^\mathrm{I}_{ij} / M^\mathrm{II}_{ij}|
\lesssim 1.5$ for all $i,j = 1,2,3$.
Obviously this solution involves a very precise tuning between the
$M_d$ and $M_\ell$ matrices such that the difference becomes close to
singular. Changing the input values for the down-quark masses and the
charged lepton mixing matrix $W_\ell$ by a factor of $(1 + 10^{-4})$
destroys in general the fit and leads to $\chi^2$ values of order
100.\footnote{The type-I, type-II, and mixed solutions require tuning
of the parameters with a typical accuracy of better than $10^{-3}$ (the
tuning is slightly less severe for the type-II case).}
Let us add that we obtain {\it mixed'} type solutions by
extrapolating from $|\xi| = 10^{-4}$ down to very small values of
$|\xi| \lesssim 10^{-9}$. However, at some point these results are
likely to become unreliable, since the numerical inversion of a
nearly singular matrix is limited by the accuracy of the algorithm.

\begin{figure}[t]
\centering
\includegraphics[width=0.48\textwidth]{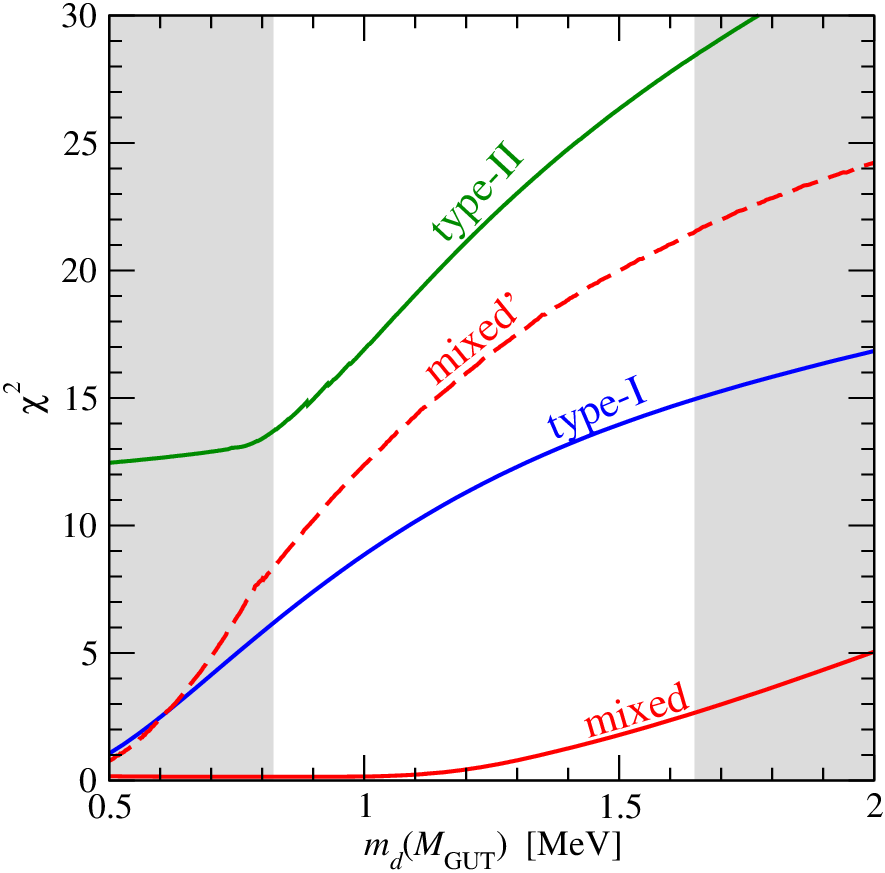}
%-eps-includegraphics[width=0.48\textwidth]{figs/md-chisq.eps}
%
  \mycaption{$\chi^2$ as a function of $m_d$ for the type-I, type-II,
  mixed, and mixed' solutions given in Tab.~\ref{tab:solutions}. The
  unshaded region corresponds to the $1\sigma$ interval for $m_d$
  from Tab.~\ref{tab:data}.}
\label{fig:md}
\end{figure}

\subsection{The role of $m_{d}$}

The down-quark mass plays a relevant role in obtaining a good fit to the data.
As it is visible from Tab.~\ref{tab:solutions} the type-II,
mixed', and type-I solutions require values of $m_d$ which are about
$1.2\sigma$, $1.8\sigma$, and $1.9\sigma$ below the preferred value
given in Tab.~\ref{tab:data}. In Fig.~\ref{fig:md} we illustrate how
the fits become worse when $m_d$ is increased. Especially the very good
fits of the type-I and mixed' solutions with $\chi^2 \simeq 4$ are
strongly affected by increasing $m_d$, and values of $m_d \simeq
1.5$~MeV at $M_\mathrm{GUT}$ lead to $\chi^2 \simeq 15$ and $20$,
respectively. Also for the pure type-II case the fit soon becomes
unacceptable if $m_d$ is increased. Only the mixed solution has enough
freedom to accomodate larger values of the down-quark mass, and in
this case also a very good fit is possible for $m_d \simeq 2$~MeV.

As a matter of fact, it is quite easy to understand these results on an analytical
basis. As it was pointed out in Refs. \cite{Goh:2003hf} and
\cite{Bertolini:2005qb} the quality of the charged sector fit is
gauged by the need to reproduce the tiny electron mass that
obeys an approximate formula of the form~\cite{Bertolini:2005qb}:
\begin{eqnarray}
|k'\,\tilde m_{e}|{\rm e}^{i \psi} & = &
 -|\tilde{r}| \,{\rm e}^{i \beta_{1}}F_{d}\lambda^{4}+{\rm e}^{i \alpha_{2}} F_{c}\lambda^{6}
 \label{meminimal}
\\
& & -A^{2}\Lambda^{2}{\rm e}^{i \alpha_{3}}\lambda^{6}
\frac{|\tilde{r}|}{{\rm e}^{i \alpha_{3}}-|\tilde{r}|}+{\cal O}(\lambda^{7})\nn
\end{eqnarray}
where $\tilde{m}_{e}\equiv
m_{e}/m_{\tau}$, $\Lambda\equiv 1-\rho-i\eta$ ($\rho$, $\eta$ being
the Wolfenstein CKM parameters and $\lambda$ the Cabibbo angle), while
$F_{d}\equiv \frac{m_{d}}{m_{b}}/\lambda^{4}$, $F_{c}\equiv
\frac{m_{c}}{m_{t}}/\lambda^{4}$ are ${\cal O}(1)$ factors.
The CKM phases $\alpha_i,\ \beta_i$ are defined in Ref.~\cite{Bertolini:2005qb}.
The parameters $\tilde{r}$ and $k'$ are given by
\begin{equation}
|k'|=\frac{m_{\tau}}{m_{t}}f_{u}|r-1|, \qquad
|\tilde{r}|=\frac{m_{b}}{m_{t}}f_{u}|r+3|.
\end{equation}
It is clear that in order to get near the physical value
$\tilde{m}_{e}\sim 2.5 \times 10^{-4}$ for $|k'|\sim 0.25$ (as
suggested by the relevant trace identities), the dominant first term
on the RHS of Eq.~(\ref{meminimal}) must be either strongly
suppressed (leading to the observed effect of low $m_{d}$ preference)
or cancelled to a large extent by the subleading ones (often at odds
with the CKM phase in the first quadrant,
c.f.~\cite{Goh:2003hf,Bertolini:2005qb}). Therefore,
for low values of $m_{d}$ the available portion of the parameter space
is larger and allows for a better global fit of the remaining physical
parameters.  It is perhaps worth mentioning that there is another
pattern in our data (though by far much weaker than the low $m_{d}$
preference) that can be justified on the same grounds, namely a
generic drift towards lower $m_{t}$ and higher $m_{c}$ regions. One verifies
that in such a case the subdominant term proportional to
$F_{c}$ gets larger and allows for a better ``screening'' of the
dominant first term on the RHS of formula (\ref{meminimal}).

\begin{figure*}[t]
\centering
\includegraphics[width=.9\textwidth]{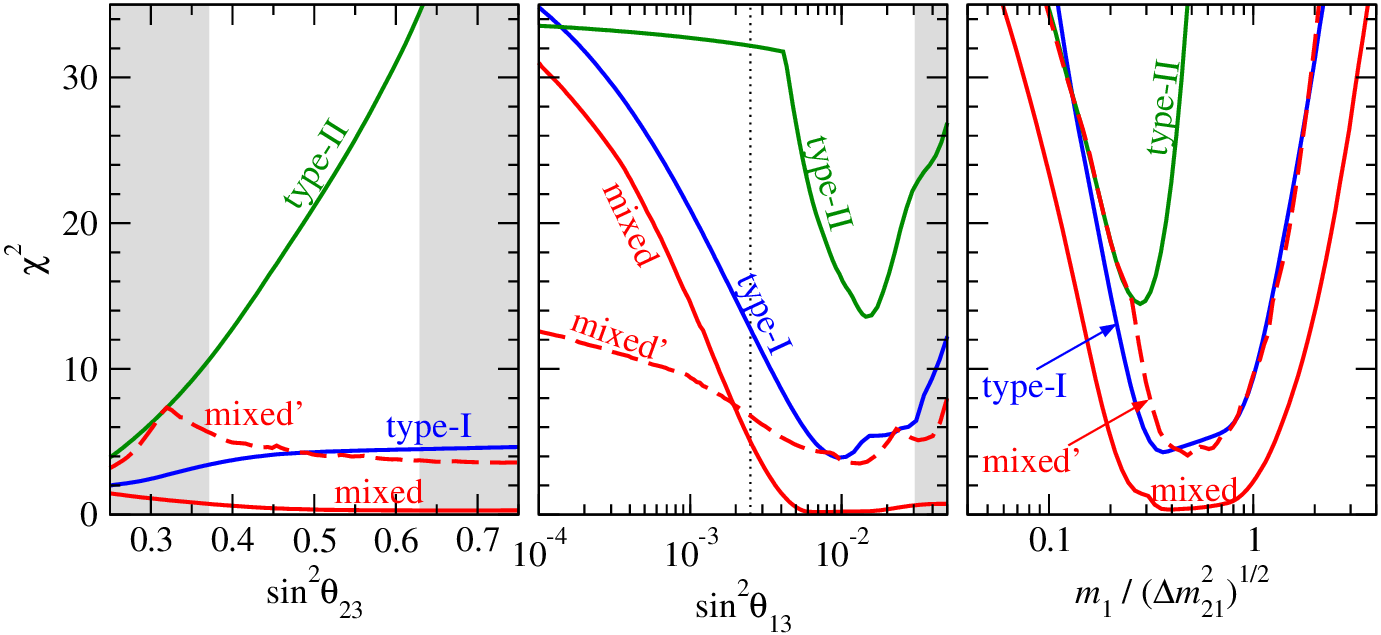}
%-eps-includegraphics[width=.9\textwidth]{figs/nu-params.eps}
%
  \mycaption{$\chi^2$ as a function of
  $\sin^2\theta_{23}^\mathrm{PMNS}$ (left),
  $\sin^2\theta_{13}^\mathrm{PMNS}$ (middle) and of $R \equiv
  m_1/\sqrt{\Delta m^2_{21}}$ (right) for the type-I, type-II, mixed,
  and mixed' solutions given in Tab.~\ref{tab:solutions}.  The shaded
  regions are excluded at $2\sigma$ according to the data given in
  Tab.~\ref{tab:data}. The dotted vertical line shows roughly the
  sensitivity to $\theta_{13}^\mathrm{PMNS}$ of neutrino oscillation
  experiments within a timescale of 10 years~\cite{Huber:2004ug}.}
\label{fig:nu-params}
\end{figure*}

\subsection{Predictions for the neutrino parameters}

%\label{sec:nu-results}

In this section we discuss in some detail the predictions of the studied setup
for the neutrino sector. Our main results are summarized in
Fig.~\ref{fig:nu-params}, where we show how the $\chi^2$ changes if
$\theta_{23}^\mathrm{PMNS}$, $\theta_{13}^\mathrm{PMNS}$, and the mass
of the lightest neutrino $m_1$ are varied. Technically this analysis is
performed in the following way: to test the variation of an observable
$O_k$ the term with $i=k$ is removed from the $\chi^2$ given in
Eq.~(\ref{eq:chisq}) (this ensures that the only constraint on the
observable comes from the mass sum rules and not from the input data).
Then, in order to test a certain value $O^*$ for the observable $O_k$ a
term is added to the $\chi^2$ with a very small error of 1\%, to
confine the fit: $(P_k(x_\alpha) - O^*)^2 / (0.01\, O^*)^2$. After the
minimization this term is removed, and the $\chi^2$ is evaluated at
the point obtained in the minimization.

Fig.~\ref{fig:nu-params} (left) shows the constraint on the
PMNS mixing angle $\theta_{23}^\mathrm{PMNS}$. One can see that for the type-I,
mixed, and mixed' solutions there is no definite prediction for this
angle and very good fits are possible with values of
$\theta_{23}^\mathrm{PMNS}$ in the whole range allowed by the data. In
contrast a scenario with pure type-II seesaw shows a clear preference
for small values of $\sin^2\theta_{23}^\mathrm{PMNS}$. In particular,
maximal mixing $\sin^2\theta_{23}^\mathrm{PMNS} = 0.5$ is disfavoured
with respect to the best fit value $\sin^2\theta_{23}^\mathrm{PMNS} =
0.36$ with $\Delta \chi^2 \approx 11$. Hence, the pure type-II seesaw
model predicts sizable deviations from maximal mixing within the reach
of upcoming neutrino oscillation experiments, see, \eg,
Refs.~\cite{Schwetz:2005jr,Antusch:2004yx}. This result is in
agreement with Refs.~\cite{Bertolini:2005qb,Babu:2005ia}. Let us
stress, however, that deviations from maximal mixing are {\it not} a
general prediction of the $SO(10)$ model under consideration; it holds
only for the pure type-II case.

Concerning the mixing angle $\theta_{13}^\mathrm{PMNS}$, for all
solutions the best fit point predicts values close to the present
upper bound and clearly within the reach of upcoming neutrino
oscillation experiments~\cite{Schwetz:2005jr,Huber:2004ug}. Also in
this case the pure type-II solution gives the most stringent
prediction. However, if the fit is stretched to some degree, the type-I
and mixed solutions allow also for smaller values of
$\theta_{13}^\mathrm{PMNS}$ that might be difficult to detect in the
next round of experiments. For instance, in the mixed solution case a fit
with $\chi^2 \approx 7$ is possible for
$\sin^2\theta_{13}^\mathrm{PMNS} = 2\times 10^{-3}$.
An interesting feature of the mixed' solution is that even for very
tiny values of $\sin^2\theta_{13}^\mathrm{PMNS} \lesssim 10^{-4}$ a
reasonable fit can be obtained with $\chi^2 \approx 12.6$. The main
contributions to the $\chi^2$ in this case are deviations of
$-1.7\sigma$ for $m_d$, $-1.2\sigma$ for $m_s$, $-2.0\sigma$ for
$m_t$, and $+1.4\sigma$ for $\sin^2\theta_{23}^\mathrm{PMNS}$. The
existence of this solution shows that if no signal of
$\theta_{13}^\mathrm{PMNS}$ is detected by the upcoming experiments it
might be still possible to construct models with viable predictions for
fermion masses and mixings, although the amount of fine tuning will
increase.

Another interesting information the setup under consideration provides, concerns
the shape of the neutrino mass spectrum. The most solid prediction is the normal
mass ordering, which means that $m_1 < m_2 < m_3$, \ie, $\Delta
m^2_{31} > 0$; no viable solution has been found for inverted ordering
($\Delta m^2_{31} < 0$) which is equally allowed by present
oscillation data.
The predictions for the absolute neutrino mass scale are given in
terms of the ratio of the lightest neutrino mass $m_1$ to the square
root of the ``solar'' mass-squared difference:
\begin{equation}\label{eq:defR}
R \equiv \frac{m_1}{\sqrt{\Delta m^2_{21}}} \,.
\end{equation}
The best fit values for this ratio given in Tab.~\ref{tab:solutions}
for the four example solutions are in the range $0.28 \le R \le
0.48$. These values show that there is only a modest hierarchy in the
neutrino masses. For example, a value of $R = 0.3$ implies that $m_2
\simeq 3.5\, m_1$, \ie, $m_1$ and $m_2$ are of the same order of
magnitude.
From the plot in the right panel of Fig.~\ref{fig:nu-params} one can
infer the allowed ranges for the ratio $R$.  We find
reasonable fits for values in the range
\begin{equation}\label{eq:R}
0.2 \lesssim R \lesssim 2 \,,
\end{equation}
where the upper bound implies $m_2 \simeq 1.1\, m_1$. For the pure
type-II solution the ratio $R$ is stronger constrained to values
around the best fit point of $0.28$, whereas the constraint is weakest
for the mixed solution.
Note that a quasi-degenerate neutrino spectrum with $m_1 \simeq
m_2 \simeq m_3$ would correspond to $R \gtrsim \sqrt{ \Delta m^2_{31}
/ \Delta m^2_{21}} \simeq 5.3$ which is clearly excluded within the
$SO(10)$ framework under consideration.
The range for $R$ given in Eq.~(\ref{eq:R}) implies the following
intervals for $m_1$ and the sum of the neutrino masses $\Sigma \equiv
m_1+m_2+m_3$:
\begin{equation}
\begin{array}{c}
1.8\times 10^{-3}\,\mathrm{eV}
\lesssim m_1 \lesssim
1.8\times 10^{-2}\,\mathrm{eV}\,,\\[1ex]
0.058\,\mathrm{eV} \lesssim \Sigma \lesssim 0.088\,\mathrm{eV} \,.
\end{array}
\end{equation}
For comparison we note that for $m_1=0$ one has $\Sigma = 0.056$~eV,
which shows that the smallest values of $m_1$, for which reasonable
fits are found, are close to the case where the contribution of $m_1$
to $\Sigma$ can be neglected.\footnote{The fact that the fit always
gives a hierarchical neutrino mass spectrum justifies a posteriori the
neglect of the running of the neutrino parameters~\cite{Antusch:2003kp,Antusch:2005gp}.}

For any given set of input parameters the values of the
phases in the PMNS matrix are determined as well. We have investigated 
in the four cases of Tab.~\ref{tab:solutions} 
the predictions for the Dirac CP phase $\delta_\mathrm{PMNS}$, as well as the two Majorana
phases $\alpha_1$ and $\alpha_2$, defined in analogy with Eq.~(\ref{eq:Vckm}) 
in the leptonic sector. 
In all cases we find a correlation among the phases, such that 
\beq
\delta_\mathrm{PMNS} \approx -2\alpha_1 \approx -2\alpha_2 + \pi. 
\label{phasecorr}
\eeq
For the mixed, mixed', and type-I solutions there
is no definite prediction for the values of the phases, and, given the
correlations above, fits of comparable quality are found for all
values of the phases in the $[0,2\pi]$ range. 
Only in the case of pure type-II
seesaw the fit prefers values of $\delta_\mathrm{PMNS} \approx 0$,
$\alpha_1 \approx  0 \,(\pi)$, $\alpha_2 \approx \pm \pi/2 $, 
in agreement with the results of Ref.~\cite{Bertolini:2005qb}
(the Majorana phases $\phi_{1,2}$ there reported are defined 
according to Ref.~\cite{Antusch:2003kp} as $-2 \alpha_{1,2}$).
Moving $\delta_\mathrm{PMNS}$ from the preferred value to $\pm 60^\circ$
increases the $\chi^2$ by 10 units.

%%%%%%%%%%%%%%%%%%%%%%%%%%%%%%%%%%%%%%%%%%%%%%%%%%%%%%%%%%%%%%%%%
\section{The Minimal SUSY $SO(10)$ GUT}
%%%%%%%%%%%%%%%%%%%%%%%%%%%%%%%%%%%%%%%%%%%%%%%%%%%%%%%%%%%%%%%%%
\label{sec:minimalso10}

Having analyzed the implications on the fermion mass fit of the $10_H$
plus $\overline{126}_H$ $SO(10)$ Yukawa sector we focus now on the study of the
so called minimal renormalizable SUSY $SO(10)$ GUT \cite{Aulakh:2003kg}. The
model is characterized (in addition to $10_H$ and $\overline{126}_H$)
by the presence of the $126_H$ and $210_H$ representations in the Higgs sector. The
$126_H$ is needed to preserve the GUT-scale D-flatness while $210_{H}$
plays the dual role of triggering the spontaneous $SO(10)$ gauge
symmetry breaking and provides the necessary mixing among the $10_H$
and $\overline{126}_H$ weak doublet components needed to achieve
(after electroweak symmetry breaking) a realistic fermion spectrum.
The tiny VEV of the left-handed
triplet component of $\overline{126}_H$ responsible for the type-II
seesaw is induced via $210_H$ couplings as well.
The model is minimal in the number of parameters, 26 altogether (soft SUSY breaking aside).
The same number is found in the minimal supersymmetric standard model (MSSM) with right-handed
neutrinos and it is much less than the number of parameters in the corresponding SUSY $SU(5)$ GUT.

\begin{figure*}
\centering
\includegraphics[width=0.85\textwidth]{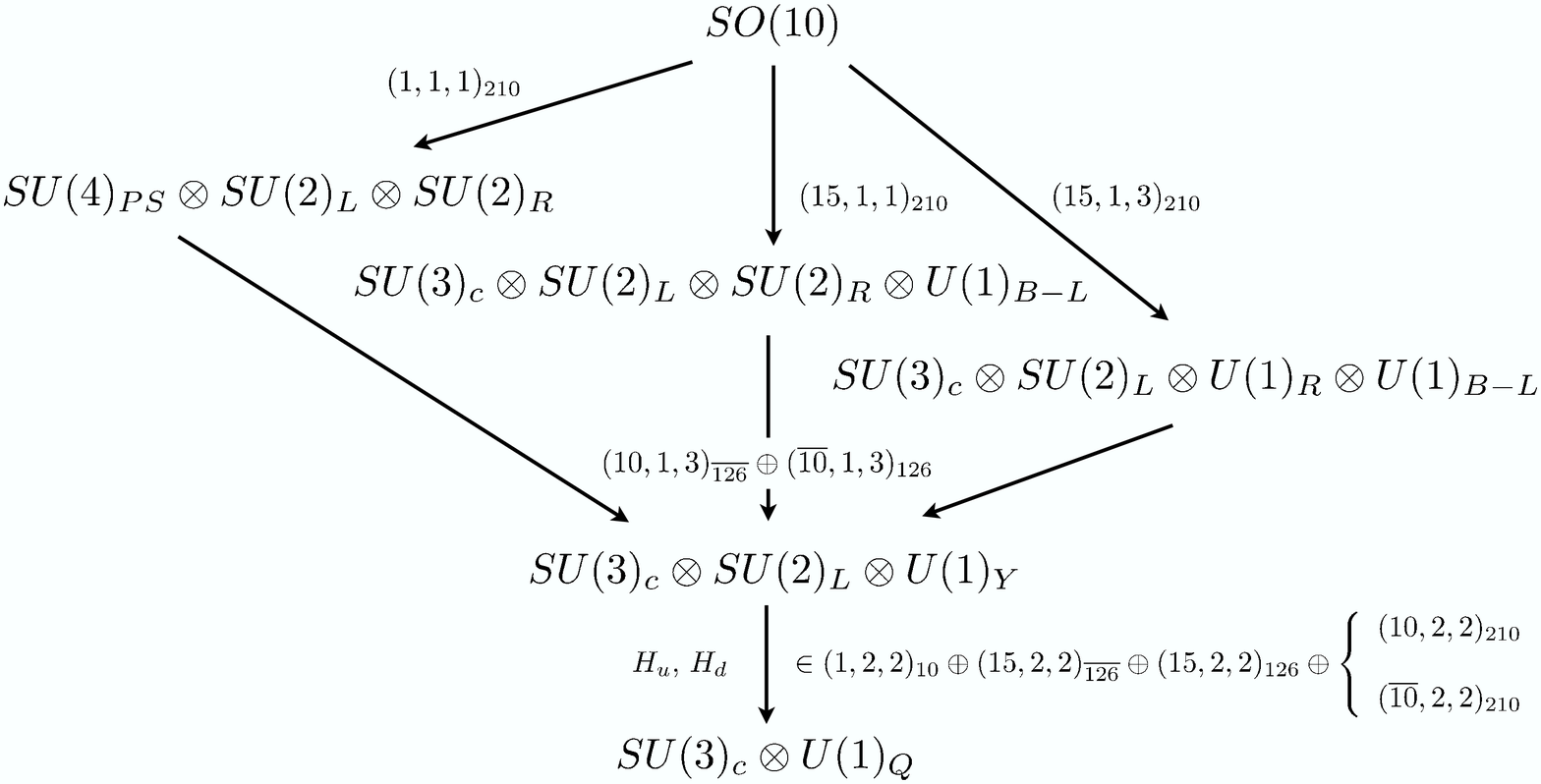}
%-eps-includegraphics[width=0.85\textwidth]{figs/mrmbreaking.eps}
%
  \mycaption{Symmetry breaking patterns and
  related VEVs in the minimal SUSY $SO(10)$. We do not display
  breaking chains involving intermediate $SU(5)$ symmetries, at
  odds with the proton decay constraints. Pati-Salam notation is used.
  }
\label{fig:mrmbreaking}
\end{figure*}

Since none of the $126_{H}$ nor $210_{H}$ Higgs multiplets can couple
to the $SO(10)$ matter bilinear $16_m \otimes 16_m$ at the
renormalizable level, the Yukawa superpotential reads
\beq
W_Y = 16_m (Y_{10} 10_H  + Y_{126} \overline{126}_H) 16_m \ ,
\label{Yukawapot}
\eeq
while the $SO(10)$ Higgs sector
is described by
\bea
W_H &=&\frac{M_{210}}{4!} {210_H}^2 + \frac{\lambda}{4!} {210_H}^3
+ \frac{M_{126}}{5!}  {126_H\ \overline{126}_H} \nn \\[1ex]
&& + \frac{\eta}{5!}  {126_H\ 210_H\ \overline{126}_H} + M_{10} {10_H}^2 \nn \\[1ex]
&& +\frac{1}{4!} {210_H\ 10_H}\ (\alpha\ {126_H}+\overline\alpha\ {\overline{126}_H})\ .
\label{Higgspot}
\eea

The minimization of the scalar potential has been analyzed in great
detail in Refs.~\cite{Bajc:2004xe,Bajc:2005qe,Aulakh:2005bd}.  The
need of a careful study of the GUT scale potential in order to perform
consistent predictions of the fermion mass textures is emphasized in
these papers.  The light MSSM-like Higgs doublets entering the low energy
Yukawa potential are in general a superposition of corresponding doublet
components of all the Higgs multiplets.  The set of VEVs which describes
the vacuum of the model is governed by one complex
parameter $x$ \cite{Bajc:2004xe}:
\bea \label{mmvevs}
&&\hspace*{-1.3em}
\vev{1,1,1}_{210} = -\frac{M_{210}}{\lambda}\
\frac{x (1-5 x^2)}{(1-x)^2}\,, \nn \\
&&\hspace*{-1.3em}
\vev{15,1,1}_{210} =
-\frac{M_{210}}{\lambda}\ \frac{(1 -2 x-x^2)}{(1-x)}\,, \label{eq:vevs} \\
&&\hspace*{-1.3em}
\vev{15,1,3}_{210} = \frac{M_{210}}{\lambda}\ x\,, \nn \\
&&\hspace*{-1.3em}
\vev{\overline{10},1,3}_{126}
\vev{10,1,3}_{\overline{126}} = \frac{2 M_{210}^2}{\eta\lambda}\
\frac{x (1-3 x) (1+x^2)}{(1-x)^2} \,.  \nn
\eea
Avoiding D-term SUSY breaking requires
$\vev{\overline{10},1,3}_{126} = \vev{10,1,3}_{\overline{126}}\equiv v_R$. In
Fig.~\ref{fig:mrmbreaking} we depict the relevant $SO(10)$ breaking
patterns together with the related VEVs.

The various mass parameters in the GUT superpotential can be written
as functions of $x$ as well. For instance
\begin{equation} M_{126} = M_{210}\ \frac{\eta}{\lambda}\
\frac{3-14 x + 15 x^2-8x^3}{(x-1)^2}\ .
\label{eq:m126}
\end{equation}
The mass parameter $M_{10}$ is determined by the minimal fine-tuning
condition~\cite{Aulakh:2003kg,Bajc:2004xe} which drives the mass of
two Higgs doublets down to the weak scale.  The relevant formula reads
\beq
M_{10} = M_{210}\ \frac{\alpha\overline\alpha}{2 \eta \lambda}\
\frac{p_{10}}{(x-1) p_3 p_5},
\label{eq:m10}
\eeq
where $p_n$ are polynomials of $x$, see Refs.~\cite{Bajc:2004xe,Bajc:2005qe}.
The $SU(2)_L$ triplet mass, relevant for the type-II seesaw, can be
written as
\beq
M_T = M_{210}\ \frac{\eta}{\lambda}\ \frac{x (4x^2-3x+1)}{(x -1)^2} ,
\label{tripletmass}
\eeq
while the tiny induced $SU(2)_{L}$ triplet VEV is given
by~\cite{Aulakh:2003kg,Aulakh:2005bd}
\beq
v_L \equiv \vev{\overline{10},3,1}_{\overline{126}} =
\frac{(\alpha v_u^{10} + \sqrt{6} \eta v_u^{\overline{126}})
v_u^{210}}{M_T} .
%= -\frac{v^2\sin^2\beta N_u^2}{M_{210}}\times \frac{\alpha^2 \lambda }{\eta \sqrt{|\eta\lambda|}}\times
%\frac{(x-1)(4 x-1) q_3^2\sigma}{x q_2}
\label{tripletvev}
\eeq
The electroweak VEV components $v_u$'s are determined in terms
of the electroweak scale and the $x$ parameter, c.f. Refs.~\cite{Bajc:2004xe,Bajc:2005qe}.

In the first part of the paper we learned that the best fit of the present data
on fermion masses and mixing is obtained by a mixed contribution of type-I
and type-II seesaws. The model independent analysis
based only on the form of the Yukawa potential in Eq.~(\ref{Yukawapot}) assumed
the vacuum state to provide the required set of parameters, in particular the correct neutrino mass scale.
In the following we examine thoroughly the question whether such a vacuum
exists in the minimal renormalizable SUSY $SO(10)$ framework. The crucial issues are
gauge coupling unification and the present limits on proton decay.
Both instances in a low energy SUSY setting privilege a great desert scenario up to
$10^{16}$~GeV. This requirement in turn tends to largely suppress
the neutrino mass scale as obtained from type-I and/or type-II seesaw.

\subsection{Matter fermion fit in the minimal renormalizable SUSY $SO(10)$ model}

For our numerical analysis we adopt the parameterization and the phase
convention of Ref.~\cite{Bajc:2005qe}. After minimal fine-tuning the
Higgs potential can be described in terms of 8 real parameters:
\bea
&&\hspace*{-1em} M_{210},\ \mbox{Re}(x),\
\mbox{Im}(x),\ \alpha,\ \overline{\alpha},\ |\lambda|,\ |\eta|, \nn\\[1ex]
&&\hspace*{-1em} \phi=\mbox{arg}(\lambda)=-\mbox{arg}(\eta)\ .
\label{higgsparameters}
\eea
For our purpose (fermion masses and symmetry breaking) without loss of
generality we can set $\phi=0$.  As emphasized in
Refs.~\cite{Bajc:2004xe,Bajc:2005qe}, the parameter $x$ provides a
systematic and effective way of describing the $SO(10)$ symmetry
breaking patterns and the related fermion mass scales.

\begin{figure*}
\centering
\includegraphics[width=0.7\textwidth]{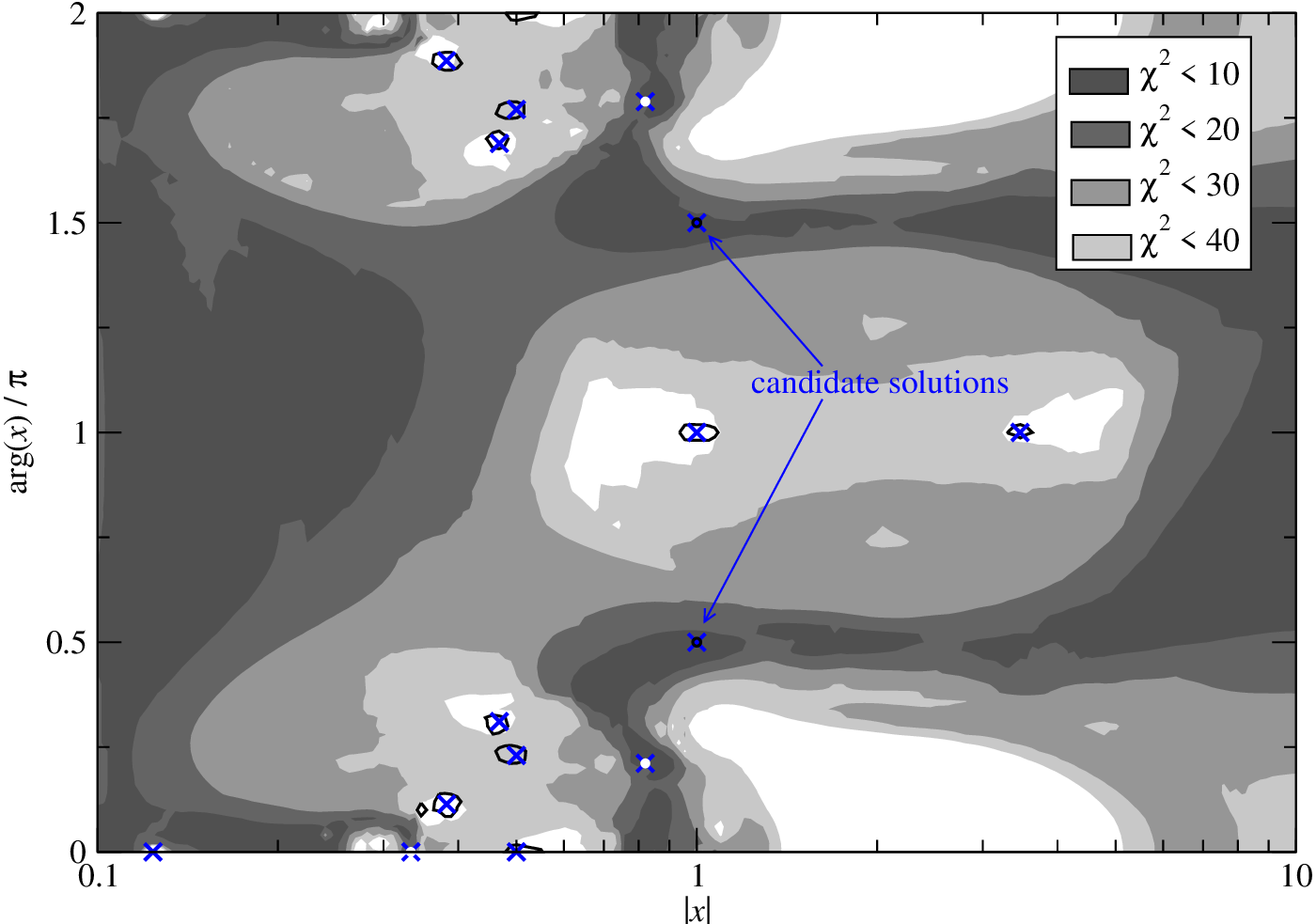}
  \mycaption{$\chi^2$ contours of the fermion mass fit in the $x$
  plane for the sample data in Table I. 
  All scalar quartic couplings are set to one.  The neutrino
  mass scale factor $f_\nu$ is treated as a free parameter.  The black
  contour curves enclose the regions where $f_\nu(x) / f_\nu({\rm
  fit}) > 0.005$, where $f_\nu(x)$ is calculated from
  Eq.~(\ref{eq:higgs-constraints}) and $f_\nu(\rm fit)$ is the value obtained from the
  fermion fit. The cross symbols corresponds to points where either
  $f_I$ or $f_{II}$ has a singular behaviour.}
\label{fig:x-scan}
\end{figure*}

In the minimal SUSY $SO(10)$ framework under consideration the VEVs appearing in
the fermion sum rules Eq.~(\ref{eq:mass-relations}) are no longer free
parameters (as assumed in the numerical analysis of
Sec.~\ref{sec:results}), but are functions of the Higgs potential
parameters, c.f.~\eq{higgsparameters}. Using the results and notation of
Ref.~\cite{Bajc:2005qe} one obtains the following expressions for the
VEV combinations appearing in the fermion fit as defined in
Eqs.~(\ref{eq:fu-r}) and (\ref{eq:fnu-xi}):
\begin{eqnarray}
f_u   &=& \frac{1}{4} \tan\beta\ \frac{N_u}{N_d}\ , \qquad
r      =  \frac{2 x p_6}{p_2 p_5} + 1\ , \nn\\
f_\nu &=& \frac{v}{M_{210}} \tan\beta \sin\beta\
          \alpha \sqrt{\frac{|\lambda|}{|\eta|}}\ \frac{N_u^2}{N_d} \ |f_{II}(x)|\ , \nn\\
\xi   &=& \frac{1}{16} \ \frac{f_I(x)}{f_{II}(x)}\ . \label{eq:higgs-constraints}
\end{eqnarray}
Here $N_u, N_d$ are functions of the parameters
in Eq. (\ref{higgsparameters}) given explicitly in the appendix of
Ref.~\cite{Bajc:2005qe}, $v = 174$~GeV is the electroweak scale, while
\begin{equation}\label{eq:fs}
f_I = \frac{2 p_2 p_5}{p_3 \sigma}\ ,\quad
f_{II} = \frac{(x-1)(4x-1) p_3 q_3^2 \sigma}{2x p_2 p_5 q_2}
\end{equation}
where
\begin{equation}\label{eq:sigma}
\sigma = \sqrt{\frac{2x(1-3x)(1+x^2)}{(1-x)^2}}\ ,
\end{equation}
and $p_n$, $q_n$ are polynomials of $x$ of order $n$ (see Tab.~I in
Ref.~\cite{Bajc:2005qe}).\footnote{Notice that our parameter $\xi$
controls the relative weight of type-I and II seesaw, and is different
from the $\xi(x)$ used in Ref.~\cite{Bajc:2005qe}.}

We now address the question of whether a realistic fermion fit is
possible, given the constraints from the Higgs sector encoded in
Eqs.~(\ref{eq:higgs-constraints}) which together with
Eqs.~(\ref{eq:Mu}), (\ref{eq:MD}), and (\ref{eq:Mnu2}) lead to a
highly constrained system of relations.
First we note that not much freedom is left in adjusting the
parameters $\alpha, \overline{\alpha}, \eta, \lambda$. This follows
from the functional form of the VEVs in Eqs.~(\ref{eq:vevs}),
of the potential mass parameters, and of $f_\nu, N_u, N_d$, as
well as from the requirement of perturbativity of the Higgs
potential. Hence, in what follows we set $\alpha \sim
\overline{\alpha}\sim \eta\sim \lambda\sim O(1)$~\cite{Bajc:2005qe}.
Similar considerations suggest also $|x|\sim O(1)$, and therefore we first restrict
our search to the range $0.1 \le |x| \le 10$ (deferring a discussion on
the small and large $x$ regimes to the end of the section).

In order to identify candidate solutions we perform a scan in
the complex $x$ plane by first fixing $\alpha = \overline{\alpha} =
\eta = \lambda = 1$. The $x$ dependence of $r$ and $\xi$ is taken into
account according to Eq.~(\ref{eq:higgs-constraints}), leaving however
the neutrino mass normalization $f_\nu$
as a free parameter to be determined from the fit. Also
$f_u$ becomes a free parameter, if $\tan\beta$ is allowed to
vary.\footnote{In the following we will consider $\tan\beta$ as a
free parameter within the range $10 \lesssim \tan\beta
\lesssim 55$. We may still keep the data in Table~\ref{tab:data} as reference GUT values
for the fermion parameters, since
changing $\tan\beta$ in the indicated range has a small
impact on the running~\cite{Das:2000uk}.}

\begin{table}[t]
\renewcommand{\arraystretch}{0.6}
\begin{tabular}{l|rr|rr}
  \hline\hline
  parameter         & \mc{best fit}         & \mc{$\lambda,\alpha,\bar\alpha$ fixed}  \\
  \hline
  $|x|$             & \mc{$1.0000002207$}      & \mc{$1.00000009837$} \\
  arg($x$)$/\pi$    & \mc{$0.4999999781$}      & \mc{$0.4999999815$} \\
  $\eta$            & \mc{$0.3191$}         & \mc{$0.3191$} \\
  $\lambda$     & \mc{$1.5642$}         & \mc{$1$} \\
  $\alpha$          & \mc{$2.9455$}         & \mc{$1$} \\
  $\bar\alpha$      & \mc{$4.0332$}         & \mc{$1$} \\
  $M_{210}$ [GeV]   & \mc{$2.000\times 10^{16}$}    & \mc{$2.000\times 10^{16}$}\\
  $\tan\beta$       & \mc{$41.15$}          & \mc{$45.88$} \\
  \hline
  $|\xi|$           & \mc{$7.476\times 10^4$}   & \mc{$1.513\times 10^5$} \\
  arg($\xi$)$/\pi$      & \mc{$-0.60492$}       & \mc{$-0.53116$} \\
  $|r|$             & \mc{$1.955$}          & \mc{$1.955$} \\
  arg($r$)$/\pi$    & \mc{$-0.2568$}        & \mc{$-0.2568$} \\
  $f_u$             & \mc{$12.28$}          & \mc{$11.47$} \\
  $f_\nu$           & \mc{$7.077\times 10^{-17}$}   & \mc{$3.745\times 10^{-17}$} \\
  $\vev{1,1,1}$ [GeV]   & \mc{$3.836\times 10^{16}$}    & \mc{$6.000\times 10^{16}$} \\
  $\vev{15,1,1}$ [GeV]  & \mc{$2.557\times 10^{16}$}    & \mc{$4.000\times 10^{16}$} \\
  $\vev{15,1,3}$ [GeV]  & \mc{$1.279\times 10^{16}$}    & \mc{$2.000\times 10^{16}$} \\
  $\vev{10,1,3}$ [GeV]  & \mc{$3.423\times 10^{13}$}    & \mc{$3.009\times 10^{13}$} \\
  $M_{PG}$ [GeV]    & \mc{$4.135\times 10^{10}$}    & \mc{$2.043\times 10^{10}$} \\
  $M_T$ [GeV]       & \mc{$8.655\times 10^{15}$}    & \mc{$1.354\times 10^{16}$} \\
  \hline
  observable & pred.\ & pull & pred.\ & pull \\
  \hline
  $m_d$ [MeV]                       &$0.3435 $&$\bf -2.2$ &$0.3373  $&$\bf -2.2$\\
  $m_s$ [MeV]                       &$27.91  $&$     1.2$ &$26.36   $&$ 0.90$\\
  $m_b$ [MeV]                       &$1065   $&$   0.038$ &$1112    $&$0.37 $\\[1ex]
  $m_u$ [MeV]                       &$0.5596 $&$   0.038$ &$0.5639  $&$0.056 $\\
  $m_c$ [MeV]                       &$212.8  $&$    0.15$ &$213.1   $&$0.16 $\\
  $m_t$ [MeV]                       &$82090  $&$   -0.21$ &$79550   $&$-0.19 $\\[1ex]
  $\sin\phi^\mathrm{CKM}_{23}$      &$0.0351 $&$  -0.039$ &$0.0350  $&$-0.065 $\\
  $\sin\phi^\mathrm{CKM}_{13}$      &$0.00325$&$    0.10$ &$0.00329     $&$0.18 $\\
  $\sin\phi^\mathrm{CKM}_{12}$      &$0.2244 $&$   0.033$ &$0.2244  $&$0.057 $\\
  $\delta_\mathrm{CKM}\, [^\circ]$  &$64.32  $&$    0.31$ &$70.10       $&$0.72 $\\[1ex]
  $\sin^2\theta^\mathrm{PMNS}_{23}$ &$0.4893 $&$   -0.16$ &$0.5002  $&$0.0035 $\\
  $\sin^2\theta^\mathrm{PMNS}_{13}$ &$0.0133 $&$    0.86$ &$0.01085     $&$0.70 $\\
  $\sin^2\theta^\mathrm{PMNS}_{12}$ &$0.3007 $&$   -0.37$ &$0.2953  $&$-0.59 $\\[1ex]
%  $\delta_\mathrm{PMNS}\ [^\circ]$ &$	     $&$   	$ & $       $ & $    	$\\
%  $\phi_{1}\, [^\circ]$ 	    &$	     $&$   	$ & $       $ & $    	$\\
%  $\phi_{2}\, [^\circ]$ 	    &$	     $&$   	$ & $       $ & $    	$\\[1ex]
  $\Delta m^2_{21}$[$10^{-5}$eV$^2$]&$7.875  $&$  -0.082$ &$7.860   $&$-0.13 $\\
  $\Delta m^2_{31}$[$10^{-3}$eV$^2$]&$2.199  $&$  -0.002$ &$2.274   $&$0.20 $\\
  $m_1 / \sqrt{\Delta m^2_{21}}$    &$0.3351 $&$	$ &$0.3289  $&$		$\\
  \hline
  $\chi^2$          &       & $7.25$ &      & $7.30$\\
  \hline\hline
\end{tabular}
  \mycaption{Parameter values and outcomes for the best fit solution, $x\simeq i$.
  The reference mass parameter $M_{210}$ is set at a typical MSSM GUT scale.
  The charged lepton masses are given in \eq{eq:ch_lept}.
  Deviations above $2\sigma$ in the fermion mass
  data are highlighted in boldface. The final $\chi^2$ is the sum of
  the squares of the numbers in the ``pull'' column.}
 \label{tab:bestfit}
\end{table}

In Fig.~\ref{fig:x-scan} we show the $\chi^2$ contours of the fermion
mass fit in the complex $x$ plane, where darker areas represent better
fits of the fermion masses and mixings according to the $\chi^2$ values
given.
As it is visible from the figure we find large regions in the $x$ plane
where a very good fit with $\chi^2 < 10$ can be obtained. The crucial
question is whether within these regions it is possible to
obtain also a consistent value of $f_\nu$. From Eq.~(\ref{eq:Mnu2})
one finds that $f_\nu$ or $f_\nu |\xi|$ has to be of order
$\sqrt{\Delta m^2_{31}} / m_b \sim 5\times 10^{-11}$ to provide a
neutrino mass scale required by oscillation data, in agreement with
the values reported in Tab.~\ref{tab:solutions}. In contrast, from
Eq.~(\ref{eq:higgs-constraints}) one finds a natural size of
\begin{equation}
f_\nu \sim \frac{v \tan\beta}{M_{210}}
\sim 5\times 10^{-13}
\left(\frac{\tan\beta}{55}\right)
\end{equation}
for $M_{210} = 2\times 10^{16}$~GeV. This estimate illustrates that
in the framework under consideration a gap of about 2 orders of
magnitude exists between the generic prediction of the neutrino masses
and the scale required by the data. We have verified this expectation
numerically and for most part of the complex $x$ plane shown in
Fig.~\ref{fig:x-scan} the value of $f_\nu(x)$ according to
Eq.~(\ref{eq:higgs-constraints}) is 2 to 4 orders of magnitude smaller
than the value of $f_\nu$ required by the fit (denoted by $f_\nu({\rm fit})$).

The only way left to obtain the needed neutrino mass scale is to consider
special points in the $x$ plane, where either $|f_{II}(x)|$
or $|f_I(x)|$ becomes large because of roots in the denominators of
Eq.~(\ref{eq:fs}). These singularities are marked in
Fig.~\ref{fig:x-scan} by the crosses. The closed black contours
correspond to the regions where $f_\nu(x) / f_\nu({\rm
fit}) > 0.005$. We find that only close to the singularities
$f_\nu$ can be large enough to provide the required neutrino mass scale.
On the other hand, it appears that most of the singularities drop into regions where
the fermion mass fit is poor.
We carefully checked
numerically the regions close to the relevant singularities and came to the
conclusion that only in the near neighborhood of $x=\pm i$ a realistic
fit of the matter sector is possible.
Helpful discussions of
specific points and limiting cases in the $x$ complex plane based on analytical considerations
can be found in Refs.~\cite{Bajc:2004xe,Bajc:2005qe}.

In Tab.~\ref{tab:bestfit} we report the best fit solution we
found. The absolute value and phase of $x$ are tuned at the level of
$10^{-7}$ in the close neighborhood of $i$. The ``symmetric'' solution $x=-i$
exhibits to quite similar features. In the ``best fit'' column we have optimized the parameters
$\alpha, \overline{\alpha}, \eta, \lambda$, whereas
in the third column we have taken $\alpha = \overline{\alpha} =
\lambda = 1$. As expected, changing
these parameters has very little impact on the fit, as it is evident from
the comparison of the second and third column. Requiring in
addition $\eta = 1$ gives a very similar result.
The neutrino mass matrix is dominated by type-I seesaw since $\sigma =
0$ for $x = \pm i$, which implies $f_I \to \infty$ and $f_{II} = 0$,
and hence $\xi \to\infty$, see Eqs.~(\ref{eq:fs}) and
(\ref{eq:higgs-constraints}). From the data in the table one finds
$f_\nu|\xi| \sim 10^{-11}$, which is the
correct value to provide the required neutrino mass scale.
The table shows that a very good fit of all quark and lepton
mass and mixing parameters (including the neutrino sector) is
obtained, with just a $2.2\sigma$ pull for $m_d$ that is acceptable in
view of the systematic uncertainities entering the light quark mass
determinations.

\subsection{Gauge couplings unification and proton decay for $x\simeq\pm i$}

What remains to be checked is whether the best fit solution satisfies unification and
proton decay constraints.
For $x=\pm i$ the
$SO(10)$ gauge symmetry is broken\footnote{We neglect the tiny
hierarchy among the VEVs of $(1,1,1)_{210}$, $(15,1,1)_{210}$, and
$(15,1,3)_{210}$ as all of them are confined for $x=\pm i$ within an interval
of less than one order of magnitude, c.f.\
Eq.~(\ref{mmvevs}) and Tab.~\ref{tab:bestfit}.} to $G_{3211} \equiv
SU(3)_C\times SU(3)_L\times U(1)_R\times U(1)_{B-L}$ by the
VEV of $(15,1,3)_{210}$ (compare Fig.~\ref{fig:mrmbreaking}
and Refs.~\cite{Bajc:2004xe,Bajc:2005qe}). The
calculation of the particle spectrum shows that in this limit a set of
unmixed states that transform as $(8,3,0,0)$ under $G_{3211}$ (or
$(8,3,0)$ under the SM gauge group) remains light.  In the exact limit
$x=\pm i$ these Goldstone states are massless because of the spontaneous breaking
of accidental global symmetries of the Higgs potential.
Upon breaking the $G_{3211}$ gauge symmetry to the SM via the VEVs
\footnote{More precisely, the quantum numbers of the states
responsible for this symmetry breaking step are:
$(1,1,\pm1,\mp2)^{3211}\in (1,1,3,\mp 2)^{3221}\in
(10,1,3)_{\overline{126}}\oplus h.c.$ }
$\vev{10,1,3}_{\overline{126}} = \vev{\overline{10},1,3}_{126} = v_R\
(\equiv M_{3211})$ they acquire a mass $M_{PG}$ of the order of
$M_{3211}^2/M_{PS}$, where $M_{PS}\equiv \vev{1,1,1}_{210}$ is the
Pati-Salam scale. According to Ref.~\cite{Bajc:2004xe} the expression
for the Pseudo-Goldstone mass is
\begin{equation}
M_{PG} = -4 M_{210} \frac{(2x -1)(x^{2}+1)}{(x-1)^2} \,,
\end{equation}
which is exactly zero for $x=\pm i$. In our best fit case reported in
Tab.~\ref{tab:bestfit} we find $M_{3211}\sim 10^{13}$~GeV, while
$M_{PS}\sim 10^{16}$~GeV. One therefore expects $M_{PG} \sim 10^{10}$
GeV, as we consistently find (see table).

\begin{figure}[t]
\centering
\includegraphics[width=0.49\textwidth]{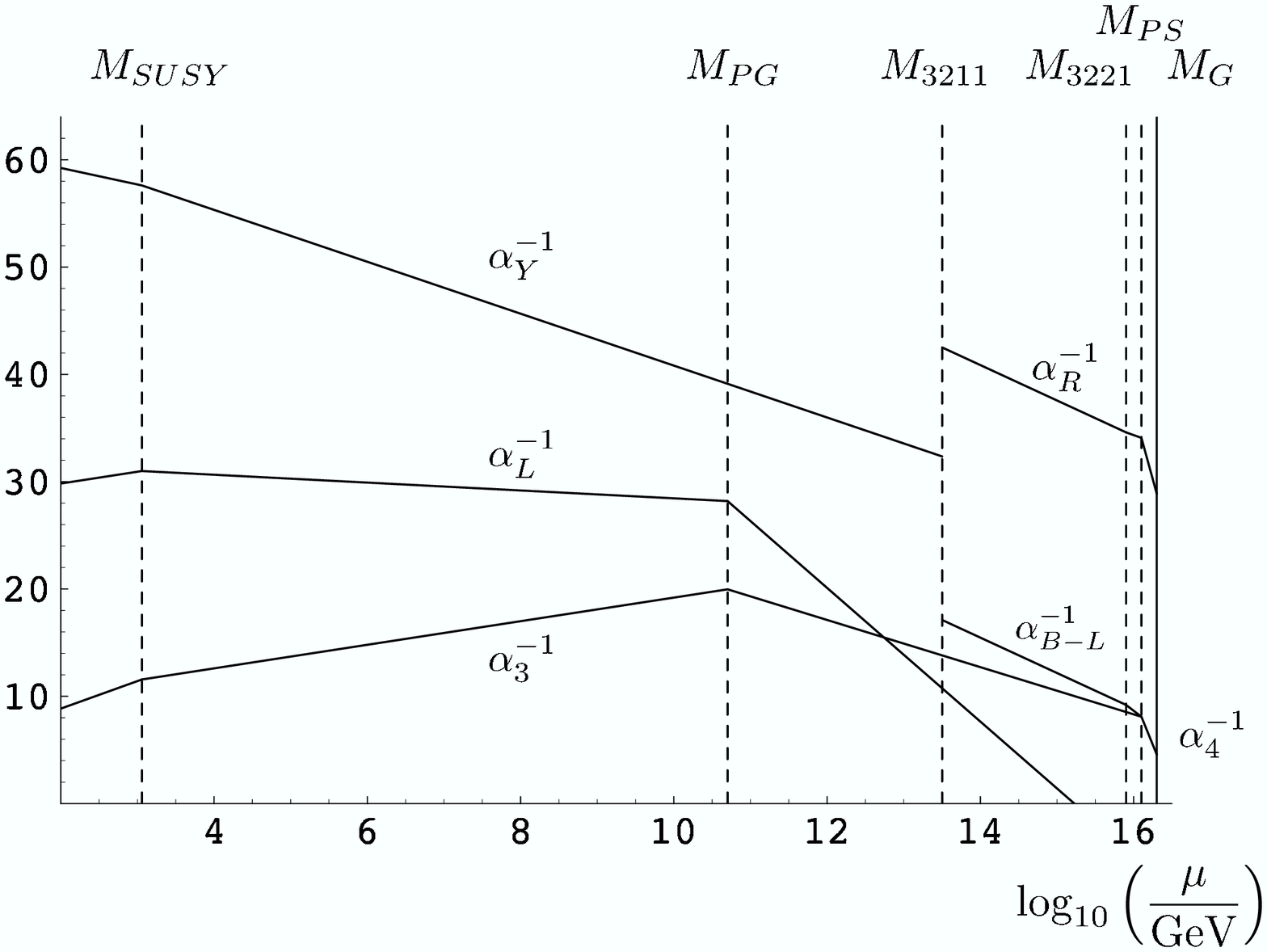}
%-eps-includegraphics[width=0.49\textwidth]{figs/running.eps}
  \mycaption{Running gauge coupling constants in the setup of
  Table~\ref{tab:bestfit}.  We have $M_\mathrm{SUSY} = 1$~TeV,
  $M_{PG}=M_{(8,3,0)}$, $M_{3211} \equiv \vev{10,1,3}$, $M_{3221}
  \equiv \vev{15,1,1}$, $M_{PS} \equiv \vev{1,1,1}$, and
  $M_\mathrm{GUT} \equiv M_{210} = 2\times 10^{16}$~GeV.}
\label{fig:running}
\end{figure}

In spite of the fact that these states are much lighter than the
GUT scale they do not affect proton decay because of their
zero $B-L$ charge.  However, they do affect heavily the running of the gauge
couplings (they transform as adjoint representations of
$SU(3)_{c}$ and $SU(2)_{L}$). Fig.~\ref{fig:running} shows how
dramatically the running is affected. The unification of the gauge
couplings is completely spoiled, with the coupling constant of
$SU(2)_L$ diverging below the GUT scale.
Hence, the only solution found providing
a realistic fit to all fermion masses and mixing parameters
has to be discarded because of the dramatic failure of the gauge couplings
unification.
The minimal renormalizable SUSY $SO(10)$ scenario seems therefore
failing to provide a realistic description of the low energy world.
In general,
if right-handed neutrinos trasform non-trivially under the GUT gauge
symmetry, the minimum neutrino mass scale required by oscillation data implies
via type I (or type II) seesaw the presence of states at
intermediate mass scales~\cite{Bajc:2005qe,Aulakh:2005mw},
that spoil the successful unification within the MSSM.

In passing let us stress that the example discussed above shows quite clearly
that the naive argument of setting all relevant particle states at the
scale of the symmetry breaking step may be far from correct.
One should always be aware that
much lighter states may appear in the spectrum as a consequence of the spontaneous
breaking of accidental (would-be) symmetries of the scalar potential~\cite{Aulakh:1982sw}.

\subsection{Large/small $|x|$ regime}

Before conclusively rejecting the minimal renormalizable SUSY $SO(10)$ scenario
we must make sure that the domain $0.1\lesssim |x|\lesssim 10$ considered in
the previous analysis covers the physically relevant region.
Indeed, the shape of the $\chi^2$ contours in
Fig.~\ref{fig:x-scan} suggests that all the matter fermion data
might be well reproduced in areas outside the considered region.
Let us give simple semianalytic arguments against this option.

By observing that in the large $x$ regime all $210_H$ VEVs in
\eq{eq:vevs} scale as $M_{210}\ |x|/\lambda$ and proton decay forces
us to maintain most of the GUT states above $M_\mathrm{GUT}\simeq
10^{16}$~GeV, we must require
\beq \frac{M_{210}\ |x|}{\lambda} \simeq 10^{16}\
\mbox{GeV}\ .
\label{large-x}
\eeq
In the large $x$ limit (and for large $\tan\beta$) from
Eqs.~(\ref{eq:higgs-constraints}), (\ref{eq:fs}) and the expressions
for $N_u, N_d$ given in Ref.~\cite{Bajc:2005qe} one finds for
the neutrino mass scales
\beq
f_\nu|\xi| \propto \frac{v}{M_{210}|x|}\, \tan\beta
\,, \
f_\nu \propto \frac{v}{M_{210}|x|}\,\frac{\tan\beta}{|x|}
\,,
\label{Mnu-large-x}
\eeq
in the type-I and type-II seesaw cases respectively.
By requiring $M_{210}|x| \simeq$~const (see Eq.~(\ref{large-x})) these
relations show that also for large $x$ the neutrino mass scale cannot
be efficiently enhanced over its natural value $v^2/M_\mathrm{GUT}$, and
accordingly our numerical calculation yields values of $f_\nu$ of
about four orders of magnitude too small in the region $|x| > 10$.
In passing let us remark that in the $|x|\gg 1$ regime there appear a
number of states with masses around $M_{210} < M_{GUT}$
that again spoil gauge unification,
albeit not affecting proton decay (see Ref.~\cite{Bajc:2004xe}).

In the small $|x|$ regime one finds for both seesaw types
\beq
f_\nu, \, f_\nu|\xi|
\propto \frac{v}{M_{210}|x|}\, \tan\beta\, \sqrt{|x|} \,.
\label{Mnu-small-x}
\eeq
Once again taking into account Eqs.~(\ref{eq:vevs},\ref{large-x}) and the fact that
scalar couplings are bound to vary in a narrow range by perturbativity
on one side and mass scale splittings on the other, no substantial
enhancement of the neutrino mass can be expected. Our $\chi^2$ fit does
confirm numerically these expectations.

We therefore conclusively show, independently confirming the
conclusions of Ref.~\cite{Aulakh:2005mw}, that the minimal renormalizable SUSY
$SO(10)$ setup in spite of noteworthy and impressive features in
reproducing the observed flavor textures fails in reproducing the
neutrino mass scale. Generally, in a renormalizable framework, type I and type II seesaws call for
intermediate scales that spoil (SUSY) gauge unification.

%%%%%%%%%%%%%%%%%%%%%%%%%%%%%%%%%%%%%%%%%%%%%%%%%%%%%%%%%%%%%%%%%%%
\section{Conclusions}
\label{sec:conclusions}

In the present paper we studied in great detail the fit of quark
and lepton masses and mixing data within a class of supersymmetric
$SO(10)$ grand unified models with a minimal renormalizable
Yukawa sector based on one $10_{H}$ and one
$\overline{126}_{H}$ Higgs representation. A systematic optimization
of the relevant $\chi^2$ function has been performed, complementary to
random parameter searches applied in previous studies. We have shown
that for a comparable size of the type-I and type-II seesaw terms,
an excellent fit to the fermion data can be obtained,
where all observables (including the CKM CP phase) are
fitted within less than $0.4$ standard deviations. Solutions
based on pure type-I and type-II seesaw have been discussed as well, and we
have identified a new class of possible solutions corresponding to a
singular behaviour of the type-II Majorana mass matrix.  The corresponding
predictions for the neutrino parameters have been investigated
in detail.

In the second part of the paper these general results were confronted
with the additional restrictions emerging in the minimal renormalizable
SUSY $SO(10)$ scenario from the model vacuum as well as from proton
decay and unification constraints. We identified a very limited
(and fine-tuned) area in the parameter space where, given the constraints from the
minimization of the Higgs potential, a fit to the quark and lepton
masses and mixing parameters is possible. However,
all solutions providing a realistic fit to the matter fermion
data had to be discarded because of the failure in reproducing the strenght of
the low energy gauge couplings.
In GUTs that embed non trivially right-handed neutrinos
the absolute neutrino mass scale required by oscillation data generally implies via type I and/or type II seesaws
the presence of states at intermediate mass scales,
that are likely to spoil the successful unification achieved in the MSSM.
Our analysis did in fact exclude, in the case of the minimal SUSY $SO(10)$ model,
the existence of (fine tuned) exceptional solutions.

The minimal renormalizable SUSY $SO(10)$ scenario based on one
$10_{H}$, one $\overline{126}_{H}$, one $126_{H}$, and one
${210}_{H}$ Higgs representations is conclusively rejected. Our
discussion suggests that either one attempts to enlarge the scalar (Yukawa) potential
in order to soften some of the parameter correlations on
the vacuum, or non-renormalizable terms are included
providing effectively the needed depletion of the seesaw scale.
Alternatively, one is let to consider the
intriguing option of a non-supersymmetric GUT scenario, where
intermediate scales, preferred by neutrino data, are needed anyway
by gauge coupling unification. Beauty and loss do strike our deepest
chords.
As the poet says:\\
{\it Is that time dead? lo! with a little rod\\
I did but touch the honey of romance---\\
And must I lose a soul's inheritance?}\\
(Oscar Wilde, {\it Helas}, 1881).

%%%%%%%%%%%%%%%%%%%%%%%%%%%%%%%%%%%%%%%%%%%%%%%%%%%%%%%%%%%%%%%%%%

\subsection*{Acknowledgments}

We thank W.~Grimus for useful discussions, C.S. Aulakh and
R.N. Mohapatra for comments on the manuscript, and T.~Shindou for
providing C++ routines used in this analysis.  The work of T.S.\ is
supported by a Marie Curie Intra-European Fellowship within the
6$^\mathrm{th}$ European Community Framework Program.  The work of
S.B. is partially supported by MIUR and the RTN European Program
MRTN-CT-2004-503369.  One of us (M.M.) is grateful to SISSA and CERN
for the hospitality during the preparation of the manuscript.

\subsection*{Note added}

The arXiv version 4 of this paper corrects three entries in Table~III,
namely the value of $\eta$ in the third column, equal to that in the second
column,
and the two values of $f_\nu$ that were off by a factor two.
We thank Borut Bajc, Ilja Dorsner and Miha Nemevsek for carefully checking
our results and spotting the typos.


\begin{thebibliography}{99}

\bibitem{GUT}
  H.~Georgi, in {\it Particles and Fields}, edited by C.~E.~Carlson
  (AIP, New York, 1975);
%
  %\bibitem{Fritzsch:1974nn}
  H.~Fritzsch and P.~Minkowski,
  {\it Unified Interactions Of Leptons And Hadrons,}
  Annals Phys.\  {\bf 93} (1975) 193.
  %%CITATION = APNYA,93,193;%%


\bibitem{seesawI}
%\bibitem{Minkowski:1977sc}
  P.~Minkowski,
  {\it Mu $\to$ E Gamma At A Rate Of One Out Of 1-Billion Muon Decays?,}
  Phys.\ Lett.\ B {\bf 67}, 421 (1977);
  %%CITATION = PHLTA,B67,421;%%
%
%\bibitem{Gell-Mann:1980vs}
  M.~Gell-Mann, P.~Ramond and R.~Slansky,
  {\it Complex Spinors And Unified Theories,}
  In {\it Supergravity}, P.~van Nieuwenhuizen and D.Z.~Freedman (eds.),
  North Holland Publ.\ Co., 1979, p.\ 315;
%Published in Stony Brook Wkshp.1979:0315 (QC178:S8:1979)
%
%\bibitem{Yanagida:1979as}
  T.~Yanagida,
  {\it Horizontal Gauge Symmetry And Masses Of Neutrinos,}
  In Proc.\ {\it Workshop on the Baryon Number of the Universe
  and Unified Theories}, O.~Sawada and A.~Sugamoto (eds.), Tsukuba, Japan,
  13--14 Feb.\ 1979, p.\ 95;
%
%\bibitem{Glashow:1979nm}
  S.L.~Glashow,
  {\it The Future Of Elementary Particle Physics,} HUTP-79-A059
  In Proc.\ Cargese 1979 {\it Quarks and Leptons}, p.\ 687;
%
%\bibitem{Mohapatra:1979ia}
  R.N.~Mohapatra and G.~Senjanovi\'c,
  {\it Neutrino Mass And Spontaneous Parity Nonconservation,}
  Phys.\ Rev.\ Lett.\  {\bf 44}, 912 (1980).
  %%CITATION = PRLTA,44,912;%%

\bibitem{seesawII}
%\bibitem{Magg:1980ut}
  M.~Magg and C.~Wetterich,
  {\it Neutrino Mass Problem And Gauge Hierarchy,}
  Phys.\ Lett.\ B {\bf 94}, 61 (1980);
  %%CITATION = PHLTA,B94,61;%%
%
%\bibitem{Lazarides:1980nt}
  G.~Lazarides, Q.~Shafi and C.~Wetterich,
  {\it Proton Lifetime And Fermion Masses In An SO(10) Model,}
  Nucl.\ Phys.\ B {\bf 181}, 287 (1981);
  %%CITATION = NUPHA,B181,287;%%
%
%\bibitem{Mohapatra:1980yp}
  R.~N.~Mohapatra and G.~Senjanovi\'c,
  {\it Neutrino Masses And Mixings In Gauge Models With Spontaneous Parity
  Violation,}
  Phys.\ Rev.\ D {\bf 23}, 165 (1981).
  %%CITATION = PHRVA,D23,165;%%

\bibitem{Aulakh:1982sw}
  C.~S.~Aulakh and R.~N.~Mohapatra,
  {\it Implications Of Supersymmetric SO(10) Grand Unification,}
  Phys.\ Rev.\ D {\bf 28}, 217 (1983).
  %%CITATION = PHRVA,D28,217;%%

\bibitem{Pati:1974yy}
  J.~C.~Pati and A.~Salam,
  {\it Lepton Number As The Fourth Color,}
  Phys.\ Rev.\ D {\bf 10} (1974) 275.
  %%CITATION = PHRVA,D10,275;%%

\bibitem{Babu:1992ia}
  K.~S.~Babu and R.~N.~Mohapatra,
  {\it Predictive neutrino spectrum in minimal SO(10) grand unification,}
  Phys.\ Rev.\ Lett.\  {\bf 70} (1993) 2845
  [hep-ph/9209215].
  %%CITATION = HEP-PH 9209215;%%

\bibitem{Lavoura:1993vz}
  L.~Lavoura,
  {\it Predicting the neutrino spectrum in minimal SO(10) grand unification,}
  Phys.\ Rev.\ D {\bf 48} (1993) 5440
  [hep-ph/9306297].
  %%CITATION = HEP-PH 9306297;%%

%%%%%%%%%%%%%%%%%%%%%%%%%%%%%%%%%%%%%%%%%%%%%%%%%%%%%%%%%%%%%%%%%%%%%%%%%%%%%%%%

\bibitem{Aulakh:2000sn}
  C.~S.~Aulakh, B.~Bajc, A.~Melfo, A.~Rasin and G.~Senjanovic,
  {\it SO(10) theory of R-parity and neutrino mass,}
  Nucl.\ Phys.\ B {\bf 597} (2001) 89
  [hep-ph/0004031].
  %%CITATION = HEP-PH 0004031;%%

\bibitem{Fukuyama:2002ch}
  T.~Fukuyama and N.~Okada,
  {\it Neutrino oscillation data versus minimal supersymmetric SO(10) model,}
  JHEP {\bf 0211} (2002) 011
  [hep-ph/0205066].
  %%CITATION = HEP-PH 0205066;%%

\bibitem{Bajc:2002iw}
  B.~Bajc, G.~Senjanovic and F.~Vissani,
  {\it b - tau unification and large atmospheric mixing:
  A case for non-canonical see-saw,}
  Phys.\ Rev.\ Lett.\  {\bf 90} (2003) 051802
  [hep-ph/0210207].
  %%CITATION = HEP-PH 0210207;%%

\bibitem{Goh:2003sy}
  H.~S.~Goh, R.~N.~Mohapatra and S.~P.~Ng,
  {\it Minimal SUSY SO(10), b tau unification and large neutrino mixings,}
  Phys.\ Lett.\ B {\bf 570}, 215 (2003)
  [hep-ph/0303055].
  %%CITATION = HEP-PH 0303055;%%

\bibitem{Fukuyama:2003hn}
  T.~Fukuyama, T.~Kikuchi and N.~Okada,
  {\it Lepton flavor violating processes and muon g-2 in minimal
  supersymmetric SO(10) model,}
  Phys.\ Rev.\ D {\bf 68} (2003) 033012
  [hep-ph/0304190];
  %%CITATION = HEP-PH 0304190;%%

\bibitem{Aulakh:2003kg}
  C.~S.~Aulakh, B.~Bajc, A.~Melfo, G.~Senjanovic and F.~Vissani,
  {\it The minimal supersymmetric grand unified theory,}
  Phys.\ Lett.\ B {\bf 588} (2004) 196
  [hep-ph/0306242].
  %%CITATION = HEP-PH 0306242;%%

\bibitem{Goh:2003hf}
  H.~S.~Goh, R.~N.~Mohapatra and S.~P.~Ng,
  {\it Minimal SUSY SO(10) model and predictions for neutrino mixings and
  leptonic CP violation,}
  Phys.\ Rev.\ D {\bf 68}, 115008 (2003)
  [hep-ph/0308197].
  %%CITATION = HEP-PH 0308197;%%

\bibitem{Bajc:2004xe}
  B.~Bajc, A.~Melfo, G.~Senjanovic and F.~Vissani,
  {\it The minimal supersymmetric grand unified theory. I: Symmetry breaking  and the particle spectrum},
  Phys.\ Rev.\ D {\bf 70}, 035007 (2004)
  [hep-ph/0402122].
  %%CITATION = HEP-PH 0402122;%%

\bibitem{Bajc:2004fj}
  B.~Bajc, G.~Senjanovic and F.~Vissani,
  {\it Probing the nature of the seesaw in renormalizable SO(10),}
  Phys.\ Rev.\ D {\bf 70} (2004) 093002
  [hep-ph/0402140].
  %%CITATION = HEP-PH 0402140;%%

\bibitem{Dutta:2004wv}
  B.~Dutta, Y.~Mimura and R.~N.~Mohapatra,
  {\it CKM CP violation in a minimal SO(10) model for neutrinos and its
  implications,}
  Phys.\ Rev.\ D {\bf 69} (2004) 115014
  [hep-ph/0402113].
  %%CITATION = HEP-PH 0402113;%%

\bibitem{Goh:2004fy}
  H.~S.~Goh, R.~N.~Mohapatra and S.~Nasri,
  {\it SO(10) symmetry breaking and type II seesaw,}
  Phys.\ Rev.\ D {\bf 70}, 075022 (2004)
  [arXiv:hep-ph/0408139].
  %%CITATION = HEP-PH 0408139;%%

\bibitem{Aulakh:2002zr}
  C.~S.~Aulakh and A.~Girdhar,
  {\it SO(10) a la Pati-Salam,}
  Int.\ J.\ Mod.\ Phys.\ A {\bf 20}, 865 (2005)
  [hep-ph/0204097];
  %%CITATION = HEP-PH 0204097;%%
%
%\bibitem{Aulakh:2004hm}
%  C.~S.~Aulakh and A.~Girdhar,
  {\it SO(10) MSGUT: spectra, couplings and thresholds effects,}
  Nucl.\ Phys.\ B {\bf 711}, 275 (2005)
  [hep-ph/0405074].
  %%CITATION = HEP-PH 0405074;%%

\bibitem{Fukuyama:2004ti}
  T.~Fukuyama, A.~Ilakovac, T.~Kikuchi, S.~Meljanac and N.~Okada,
  {\it Higgs masses in the minimal SUSY SO(10) GUT,}
  Phys.\ Rev.\ D {\bf 72} (2005) 051701
  [hep-ph/0412348].
  %%CITATION = HEP-PH 0412348;%%

\bibitem{Aulakh:2005ic}
  C.~S.~Aulakh,
  {\it Consistency of the minimal supersymmetric GUT spectra,}
  Phys.\ Rev.\ D {\bf 72}, 051702 (2005).
  %%CITATION = PHRVA,D72,051702;%%

\bibitem{Fukuyama:2005us}
  T.~Fukuyama, T.~Kikuchi and T.~Osaka,
  {\it Non-thermal leptogenesis and a prediction of inflaton mass in a
  supersymmetric SO(10) model,}
  JCAP {\bf 0506} (2005) 005
  [hep-ph/0503201].
  %%CITATION = HEP-PH 0503201;%%

\bibitem{Bertolini:2004eq}
  S.~Bertolini, M.~Frigerio and M.~Malinsky,
  {\it Fermion masses in a SUSY SO(10) model with type II seesaw: A non-minimal
  predictive scenario,}
  Phys.\ Rev.\ D {\bf 70}, 095002 (2004)
  [hep-ph/0406117].
  %%CITATION = HEP-PH 0406117;%%

\bibitem{Bertolini:2005qb}
  S.~Bertolini and M.~Malinsky,
  {\it On CP violation in a minimal renormalizable SUSY SO(10) model and beyond},
  Phys.\ Rev.\ D {\bf 72}, 055021 (2005)
  [hep-ph/0504241].
  %%CITATION = HEP-PH 0504241;%%

\bibitem{Babu:2005ia}
  K.~S.~Babu and C.~Macesanu,
  {\it Neutrino masses and mixings in a minimal SO(10) model,}
  Phys.\ Rev.\ D {\bf 72} (2005) 115003
  [hep-ph/0505200].
  %%CITATION = HEP-PH 0505200;%%

\bibitem{Dutta:2005ni}
  B.~Dutta, Y.~Mimura and R.~N.~Mohapatra,
  {\it Neutrino mixing predictions of a minimal SO(10) model with
  suppressed proton decay,}
  Phys.\ Rev.\ D {\bf 72}, 075009 (2005)
  [hep-ph/0507319].
  %%CITATION = HEP-PH 0507319;%%

%%%%%%%%%%%%%%%%%%%%%%%%%%%%%%%%%%%%%%%%%%%%%
%   proton decay
%%%%%%%%%%%%%%%%%%%%%%%%%%%%%%%%%%%%%%%%%%%%%

\bibitem{Goh:2003nv}
  H.~S.~Goh, R.~N.~Mohapatra, S.~Nasri and S.~P.~Ng,
  {\it Proton decay in a minimal SUSY SO(10) model for neutrino mixings,}
  Phys.\ Lett.\ B {\bf 587}, 105 (2004)
  [hep-ph/0311330].
  %%CITATION = HEP-PH 0311330;%%

\bibitem{Fukuyama:2004pb}
  T.~Fukuyama, A.~Ilakovac, T.~Kikuchi, S.~Meljanac and N.~Okada,
  {\it Detailed analysis of proton decay rate in the minimal
  supersymmetric  SO(10) model,}
  JHEP {\bf 0409}, 052 (2004)
  [hep-ph/0406068];
  %%CITATION = HEP-PH 0406068;%%
%
%\bibitem{Fukuyama:2004xs}
%  T.~Fukuyama, A.~Ilakovac, T.~Kikuchi, S.~Meljanac and N.~Okada,
  {\it General formulation for proton decay rate in minimal supersymmetric
  SO(10) GUT,}
  Eur.\ Phys.\ J.\ C {\bf 42} (2005) 191
  [hep-ph/0401213].
  %%CITATION = HEP-PH 0401213;%%

%%%%%%%%%%%%%%%%%%%%%%%%%%%%%%%%%%%%%%%%%%%%%%%%%%%%%%%%%%%%%%%%%%%

\bibitem{Bajc:2005qe}
  B.~Bajc, A.~Melfo, G.~Senjanovic and F.~Vissani,
  {\it Fermion mass relations and the structure of the light Higgs in a
  supersymmetric SO(10) theory,}
  Phys.\ Lett.\ B, {\bf 634} (2006) 272
  [hep-ph/0511352].
  %%CITATION = HEP-PH 0511352;%%

\bibitem{Aulakh:2005mw}
  C.~S.~Aulakh and S.~K.~Garg,
  {\it MSGUT: From Bloom to Doom,}
  [hep-ph/0512224].
  %%CITATION = HEP-PH 0512224;%%

\bibitem{Aulakh:2005bd}
  C.~S.~Aulakh,
  {\it MSGUTs from germ to bloom: Towards falsifiability and beyond,}
  [hep-ph/0506291].
  %%CITATION = HEP-PH 0506291;%%

\bibitem{Aulakh:2006vi}
  C.~S.~Aulakh,
  {\it Fermion mass hierarchy in the Nu MSGUT. I: The real core},
  [hep-ph/0602132].
  %%CITATION = HEP-PH 0602132;%%

\bibitem{Lavoura:2006dv}
  L.~Lavoura, H.~K\"uhb\"ock and W.~Grimus,
  {\it Charged-fermion masses in SO(10): Analysis with scalars in 10+120},
  [hep-ph/0603259].
  %%CITATION = HEP-PH 0603259;%%

\bibitem{Mohapatra:2006gs}
  R.~N.~Mohapatra and A.~Y.~Smirnov,
  {\it Neutrino mass and new physics},
  [hep-ph/0603118].
  %%CITATION = HEP-PH 0603118;%%

%%%%%%%%%%%%%%%%%%%%%%%%%%%%%%%%%%%%%%%%%%%%%%%%%%%%%%%%%%%%%%%%%%%

\bibitem{Das:2000uk}
  C.~R.~Das and M.~K.~Parida,
  {\it New formulas and
  predictions for running fermion masses at higher scales in SM,
  2HDM, and MSSM},
  Eur.\ Phys.\ J.\ C {\bf 20} (2001) 121
  [hep-ph/0010004].
  %%CITATION = HEP-PH 0010004;%%

\bibitem{PDG}
  Particle Data Group,
  S.~Eidelman {\it et al.}, Phys.\ Lett.\ B {\bf 592} (2004) 1
  and 2005 partial update for the 2006 edition available on
  http://pdg.lbl.gov/.

\bibitem{Antusch:2003kp}
  S.~Antusch, J.~Kersten, M.~Lindner and M.~Ratz,
  {\it Running neutrino masses, mixings and CP phases: Analytical results and
  phenomenological consequences},
  Nucl.\ Phys.\ B {\bf 674}, 401 (2003)
  [hep-ph/0305273].
  %%CITATION = HEP-PH 0305273;%%

\bibitem{Antusch:2005gp}
  S.~Antusch, J.~Kersten, M.~Lindner, M.~Ratz and M.~A.~Schmidt,
  {\it Running neutrino mass parameters in see-saw scenarios},
  JHEP {\bf 0503}, 024 (2005)
  [hep-ph/0501272].
  %%CITATION = HEP-PH 0501272;%%

\bibitem{Maltoni:2004ei}
  M.~Maltoni, T.~Schwetz, M.~A.~Tortola and J.~W.~F.~Valle,
  {\it Status of global fits to neutrino oscillations,}
  New J.\ Phys.\  {\bf 6} (2004) 122
  [hep-ph/0405172];
  %%CITATION = HEP-PH 0405172;%%

\bibitem{Schwetz:2005jr}
  T.~Schwetz,
  {\it Neutrino oscillations: Current status and prospects,}
  Acta Phys.\ Polon.\ B {\bf 36} (2005) 3203
  [hep-ph/0510331].
  %%CITATION = HEP-PH 0510331;%%

\bibitem{NR}
  See for instance, W.~H.~Press {\it et al.,}
  {\it Numerical Recipes in C: The Art of Scientific Computing},
  Cambridge University Press 1992.

\bibitem{Antusch:2004yx}
  S.~Antusch, P.~Huber, J.~Kersten, T.~Schwetz and W.~Winter,
  {\it Is there maximal mixing in the lepton sector?},
  Phys.\ Rev.\ D {\bf 70} (2004) 097302
  [hep-ph/0404268].
  %%CITATION = HEP-PH 0404268;%%

\bibitem{Huber:2004ug}
  P.~Huber, M.~Lindner, M.~Rolinec, T.~Schwetz and W.~Winter,
  {\it Prospects of accelerator and reactor neutrino oscillation
  experiments  for the coming ten years,}
  Phys.\ Rev.\ D {\bf 70} (2004) 073014
  [hep-ph/0403068].
  %%CITATION = HEP-PH 0403068;%%

\end{thebibliography}
\end{document}